\documentclass[letterpaper,english,reprint,aps,longbibliography]{revtex4-2}
\usepackage[T1]{fontenc}
\usepackage[utf8]{inputenc}
\usepackage{color}
\usepackage{babel}
\usepackage{amsmath}
\usepackage{amssymb}
\usepackage{graphicx}
\usepackage[pdfusetitle,
 bookmarks=true,bookmarksnumbered=false,bookmarksopen=false,
 breaklinks=true,pdfborder={0 0 1},backref=false,colorlinks=true]
 {hyperref}



\begin{document}
\title{Suppression of Decoherence at Exceptional Transitions}
\author{Mei-Lin Li}
\thanks{These two authors contributed equally to this work.}
\affiliation{Institute for Theoretical Physics, School of Physics, South China
Normal University, Guangzhou 510006, China}
\affiliation{Key Laboratory of Atomic and Subatomic Structure and Quantum
Control (Ministry of Education), Guangdong Basic Research Center
of Excellence for Structure and Fundamental Interactions of Matter,
School of Physics, South China Normal University, Guangzhou 510006,
China}
\affiliation{Guangdong Provincial Key Laboratory of Quantum Engineering and
Quantum Materials, Guangdong-Hong Kong Joint Laboratory of Quantum
Matter, South China Normal University, Guangzhou 510006, China}
\author{Zuo Wang}
\thanks{These two authors contributed equally to this work.}
\affiliation{Institute for Theoretical Physics, School of Physics, South China
Normal University, Guangzhou 510006, China}
\affiliation{Key Laboratory of Atomic and Subatomic Structure and Quantum
Control (Ministry of Education), Guangdong Basic Research Center
of Excellence for Structure and Fundamental Interactions of Matter,
School of Physics, South China Normal University, Guangzhou 510006,
China}
\affiliation{Guangdong Provincial Key Laboratory of Quantum Engineering and
Quantum Materials, Guangdong-Hong Kong Joint Laboratory of Quantum
Matter, South China Normal University, Guangzhou 510006, China}
\author{Liang He}
\email{liang.he@scnu.edu.cn}

\affiliation{Institute for Theoretical Physics, School of Physics, South China
Normal University, Guangzhou 510006, China}
\affiliation{Key Laboratory of Atomic and Subatomic Structure and Quantum
Control (Ministry of Education), Guangdong Basic Research Center
of Excellence for Structure and Fundamental Interactions of Matter,
School of Physics, South China Normal University, Guangzhou 510006,
China}
\affiliation{Guangdong Provincial Key Laboratory of Quantum Engineering and
Quantum Materials, Guangdong-Hong Kong Joint Laboratory of Quantum
Matter, South China Normal University, Guangzhou 510006, China}
\begin{abstract}
Decoherence is strongly influenced by environmental criticality,
with conventional Hermitian critical points typically enhancing
the loss of quantum coherence. Here, we show that this paradigm
is fundamentally altered in non-Hermitian environments. Focusing
on qubits coupled to non-Hermitian spin chains and interacting
ultracold Fermi gases, we find that approaching exceptional points
can either enhance or strongly suppress decoherence, depending
on the balance between Hermitian and non-Hermitian system--environment
couplings. In particular, when these couplings are comparable,
decoherence is dramatically suppressed at exceptional transitions.
We trace this behavior to the distinct response of the environmental
ground state near non-Hermitian degeneracies and demonstrate
the robustness of this effect across multiple models. Finally,
we show that the predicted suppression of decoherence is directly
observable on current digital quantum simulation platforms. Our
results establish exceptional points as a concrete mechanism
for suppressing decoherence and identify non-Hermitian criticality
as a new avenue for coherence control in open quantum systems
and quantum technologies. 
\end{abstract}
\maketitle
\emph{Introduction}.---The coherent superposition of quantum
states---quantum coherence---marks one of the most fundamental
distinctions between quantum mechanics and classical physics.
It underlies entanglement, which plays an essential role in quantum
computation and quantum information science \cite{Nielsen_Chuang_Book_2010}.
Yet, in realistic settings, quantum systems may often lose coherence
through interactions with their environments, a process known
as decoherence \cite{Zurek2003_RMP,Schlosshauer_RMP_2005}, thereby
becoming effectively classical. Understanding decoherence is
therefore of fundamental importance to the foundations of quantum
physics \cite{Zurek2003_RMP,Schlosshauer_RMP_2005}, as well
as practical applications such as quantum engineering \cite{Nielsen_Chuang_Book_2010}.

While environments are traditionally modeled as closed systems
in this context \cite{Caldeira_PRL_1981,Leggett_RMP_1987,Zurek2003_RMP,Schlosshauer_RMP_2005,Breuer_2007_OUP,Breuer_2016_RMP},
recent experimental advances have enabled the controlled engineering
of tailored dissipative environments \cite{Harrington_Nat_Rev_Phys_2022}.
A it has been shown that dissipative environments can sometimes
give rise to unconventional decoherence dynamics, including enhanced
qutum coherence in the presence of strong dissipation \cite{Onizhuk_2024_PRL}.
This motivates us to revisit a prototypical problem of decoherence---namely,
decoherence in the presence of environmental criticality \cite{Davide_2021_PR}---in
the new scenario.

Among dissipative environments, a case in point is non-Hermitian
environments, i.e., those described by non-Hermitian Hamiltonians
\cite{ashida2020_AP}. A hallmark feature of such environments
is the existence of exceptional transitions \cite{Fruchart_Nature_2021,ashida2020_AP},
which correspond to exceptional points (EPs)---non-Hermitian
degeneracies where two or more eigenvalues and eigenvectors coalesce
\cite{ashida2020_AP}. Much like conventional quantum critical
points, EPs can signal spectral gap closings and continuous phase
transitions, as exemplified in the Lee--Yang model \cite{G_1991_JPA,Castro_2009_JPA,Uzelac_1979_PRL}.
In conventional closed environments, decoherence is significantly
enhanced near quantum critical points, as shown theoretically
\cite{QuanHT2006_PRL,Rossini_2007_PRA,YuanZi-Gang_2007_PRA,Yuan_2007_PRA,Damski_2011_PRA,Vicari_2018_PRA,Rossini_2019_PRA}
and experimentally \cite{Zhang2008_PRL}. By contrast, dissipative
environments can in some cases enhance coherence \cite{Onizhuk_2024_PRL}.
This contrast naturally raises the fundamental question of the
fate of decoherence at exceptional transitions in non-Hermitian
environments.

\begin{figure}
\includegraphics[width=1.7in]{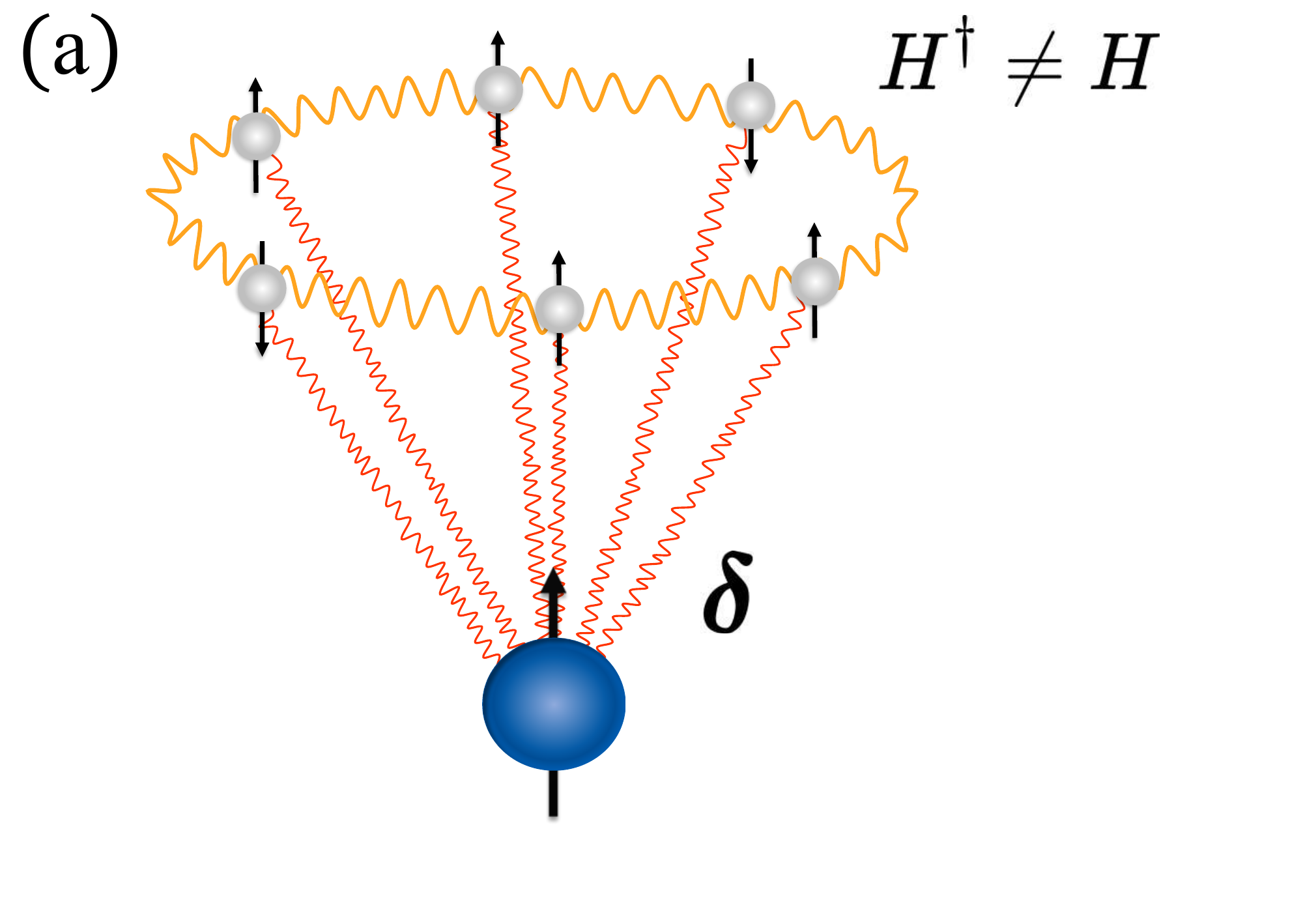}

\includegraphics[width=1.7in]{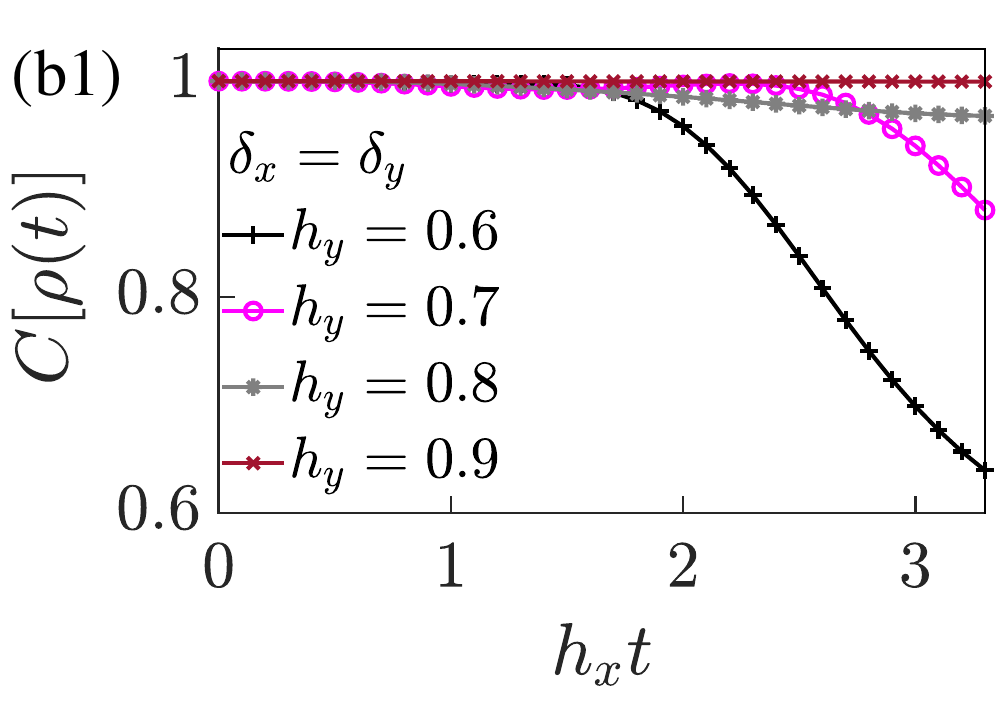}\includegraphics[width=1.7in]{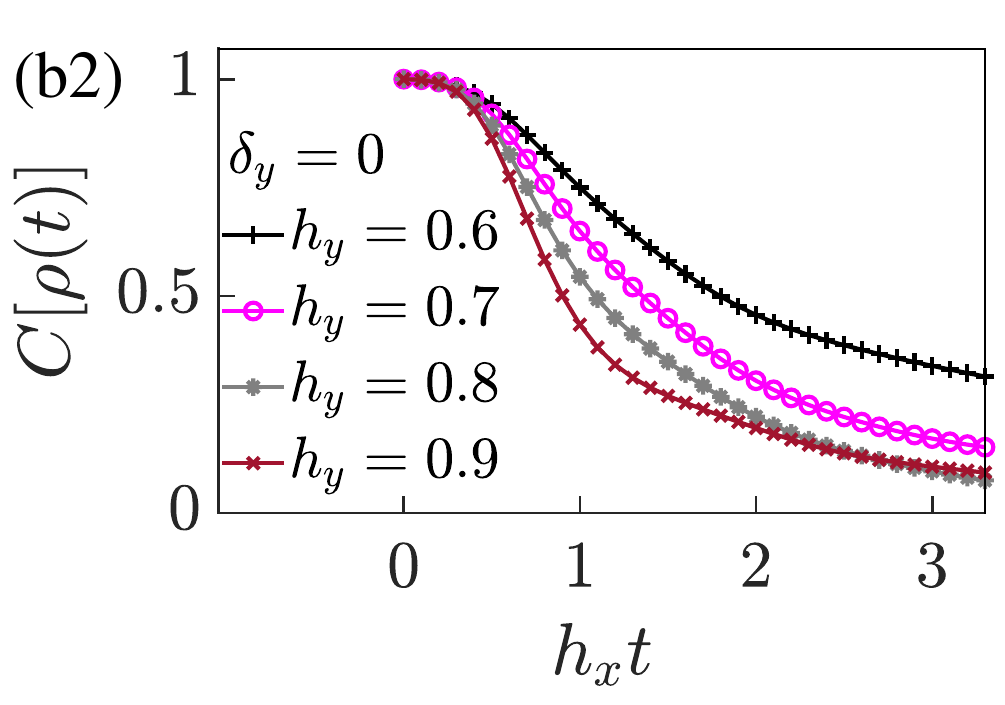}

\includegraphics[width=1.7in]{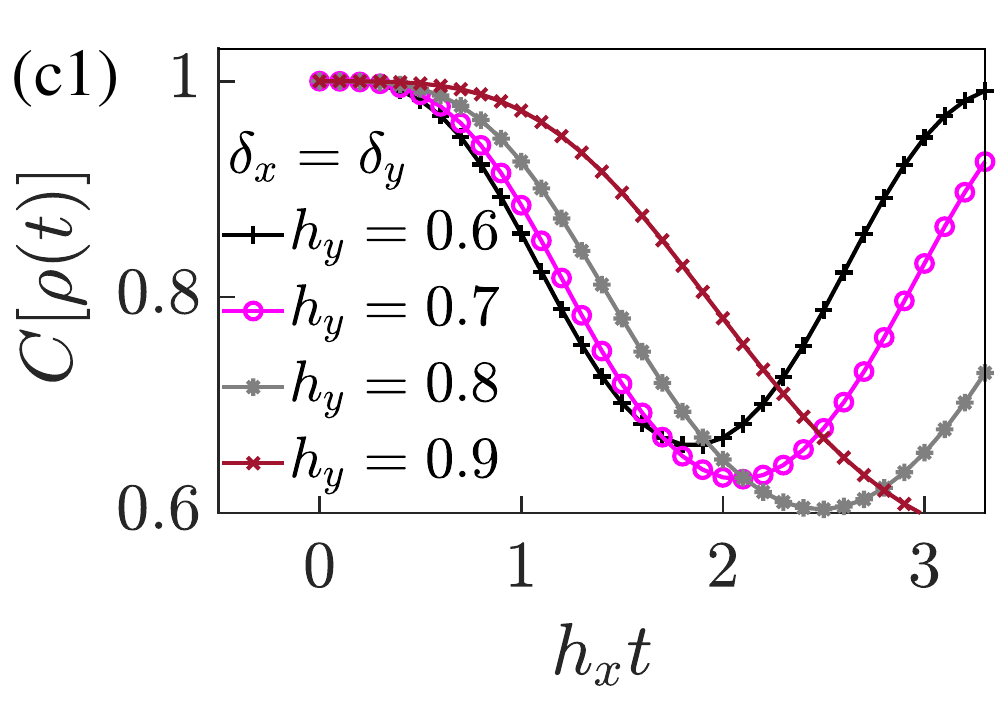}\includegraphics[width=1.7in]{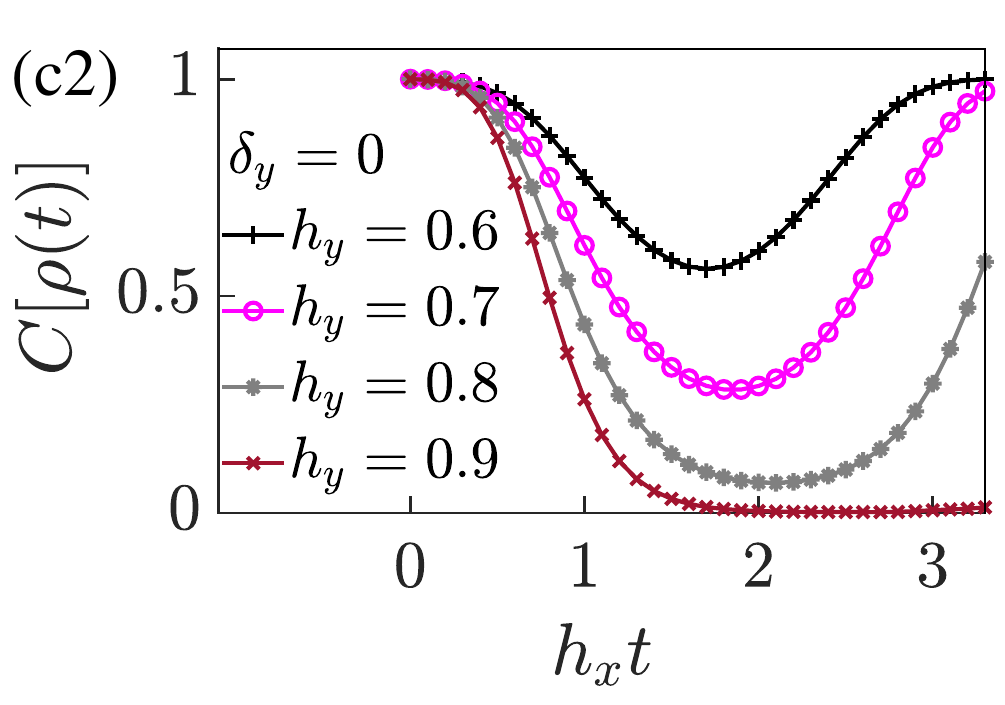}\caption{(a) Schematic illustration of a central qubit (system) coupled
to its non-Hermitian environment formed by interacting spins
in a complex transverse field. (b1, b2) Dynamics of qubit coherence
$C[\rho(t)]$ as the non-Hermitian environment approaches its
EPs, $h_{y}\rightarrow h_{x}=1$. The environment is formed by
an Ising chain with a complex transverse field {[}see Eq.~(\ref{eq:Ising_model}){]}.
(b1) Coherence dynamics for balanced Hermitian and non-Hermitian
system-environment couplings ($\delta_{x}=\delta_{y}$), showing
pronounced suppression of decoherence near the EP. (b2) Coherence
dynamics in the absence of non-Hermitian coupling ($\delta_{y}=0$),
showing the conventional enhancement of decoherence near the
EP. (c1, c2) Same as (b1, b2), but for a non-Hermitian environment
formed by Heisenberg spin chain in a complex transverse field.
All simulations use $N=25$, $J=1/2$, $h_{x}=1$, and $\sqrt{\delta_{x}^{2}+\delta_{y}^{2}}=0.1$.
See text for more details.}

\label{fig: non-Hermitian_Spin_Chains}
\end{figure}

\begin{figure}
\includegraphics[width=1.7in]{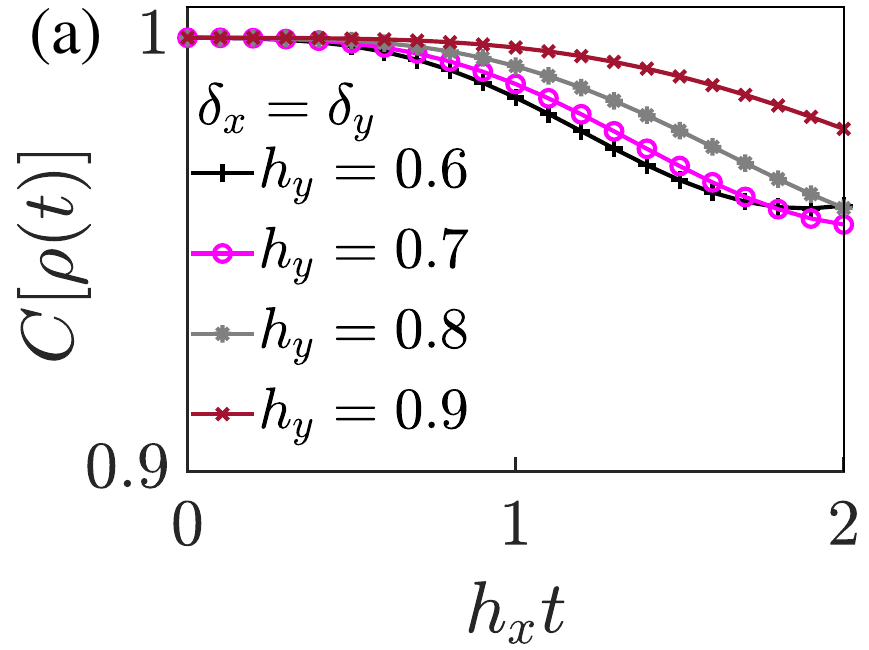}\includegraphics[width=1.7in]{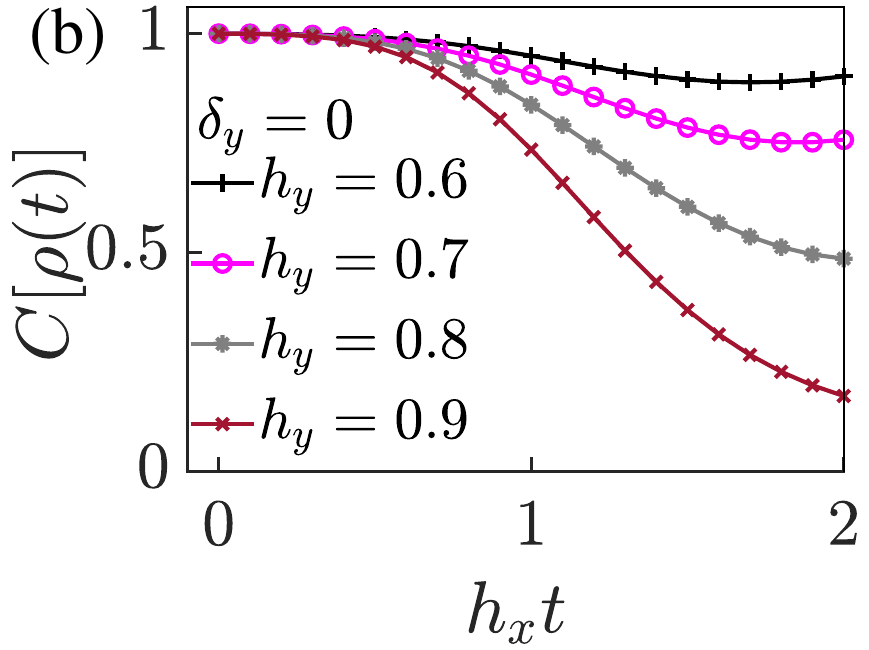}
\caption{Dynamics of qubit coherence $C[\rho(t)]$ as a non-Hermitian
environment formed by Fermi gases in optical lattices approaches
its EPs, $h_{y}\rightarrow h_{x}=1$. (a) Coherence dynamics
for balanced Hermitian and non-Hermitian system-environment couplings
($\delta_{x}=\delta_{y}$), showing pronounced suppression of
decoherence near the EP. (b) Coherence dynamics in the absence
of non-Hermitian coupling ($\delta_{y}=0$), showing the conventional
enhancement of decoherence near the EP. All simulations use $N=10$,
$J=0.1$, $U=0.4$, $h_{x}=1$, $\sqrt{\delta_{x}^{2}+\delta_{y}^{2}}=0.1$.
See text for more details.}
\label{fig: non-Hermitian_fermi gases}
\end{figure}

\begin{figure}
\includegraphics[width=3in]{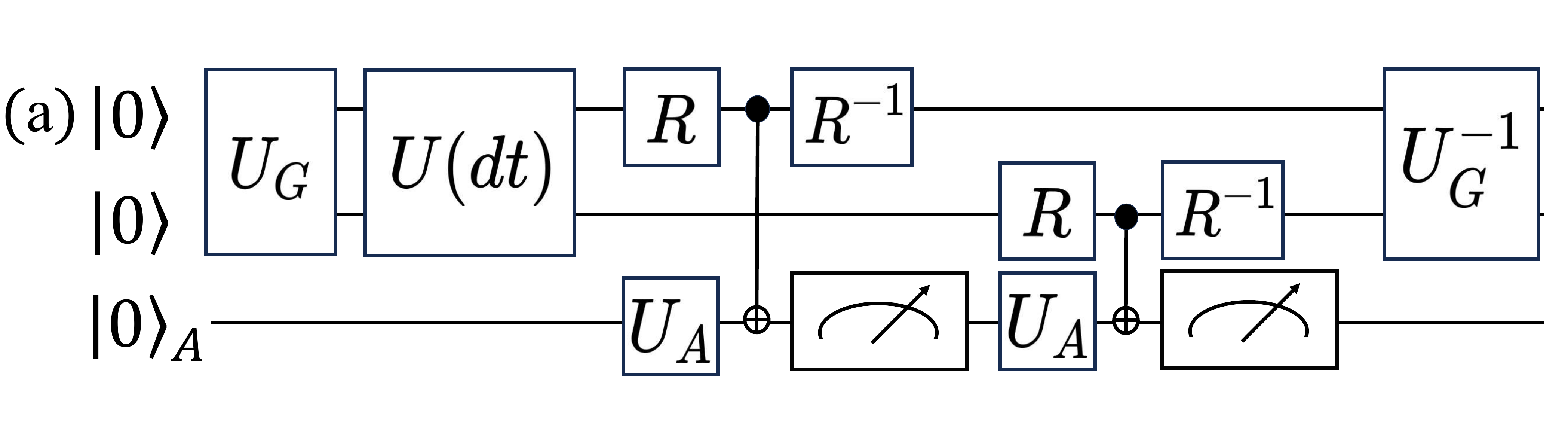}

\includegraphics[width=2in]{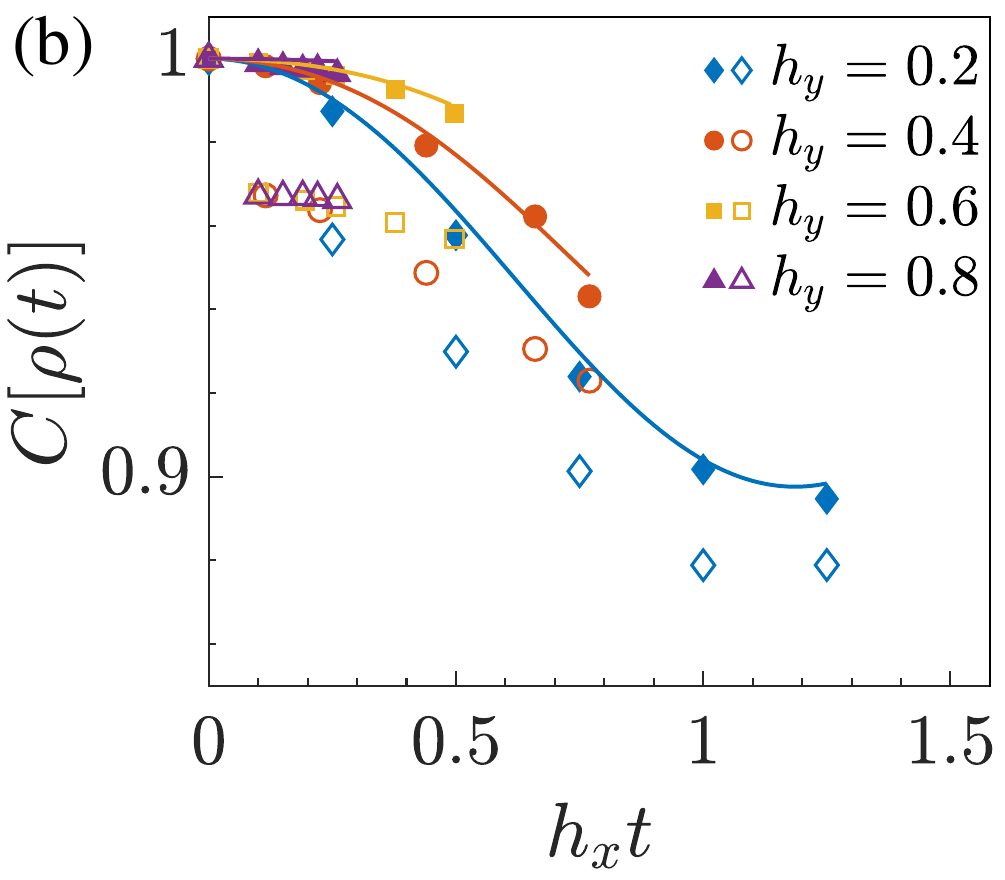}

\caption{Simulation of qubit coherence dynamics on a quantum circuit,
with the non-Hermitian environment formed by an Ising chain in
a complex transverse field. (a) Quantum circuit protocol for
simulating the first time step of the dynamics. (b) Coherence
dynamics obtained from the quantum circuit. Hollow (solid) markers
denote results from $2\times10^{6}$ trajectories on the \emph{qasm\_simulator}
with (without) noise from \emph{ibm\_brisbane}, provided by IBM
Qiskit \cite{cross_2018_ibm}. Solid curves show exact results
from numerical calculations. Simulations use $J=1/100$, $h_{x}=1$,
$\delta_{x}=\delta_{y}=0.5$, and an optimized time step $dt$.
See text for more details.}
\label{fig:IMBQ}
\end{figure}

In this work, we address this question across diverse non-Hermitian
environments, including spin chains and interacting ultracold
Fermi gases, and find that decoherence induced by non-Hermitian
environments exhibits behavior fundamentally distinct from that
near Hermitian quantum critical points (see Fig.~\ref{fig: non-Hermitian_Spin_Chains}).
More specifically, we find approaching EPs of non-Hermitian environments
can either enhance {[}see Figs.~\ref{fig: non-Hermitian_Spin_Chains}(b2,~c2),~\ref{fig: non-Hermitian_fermi gases}(b){]}
or strongly suppress decoherence {[}see Figs.~\ref{fig: non-Hermitian_Spin_Chains}(b1,~c1),~\ref{fig: non-Hermitian_fermi gases}(a){]},
depending on the balance between Hermitian and non-Hermitian
system--environment couplings. This behavior can be traced back
to the distinct responses of the environmental ground state to
different couplings. We further demonstrate the experimental
observability of the decoherence suppression at exceptional transitions
on current digital quantum computing platforms (see Fig.~\ref{fig:IMBQ}).

\emph{System and model.}---We consider a central two-level system
(qubit) coupled to a spin-chain environment, a setup widely used
in decoherence studies \cite{QuanHT2006_PRL,Rossini_2019_PRA,Rossini_2007_PRA,Yuan_2007_PRA,YuanZi-Gang_2007_PRA,Damski_2011_PRA,Vicari_2018_PRA}
(see Fig.~\ref{fig: non-Hermitian_Spin_Chains}). In Hermitian
environments, decoherence is strongly influenced by quantum criticality,
where critical fluctuations enhance the loss of coherence of
the qubit \cite{QuanHT2006_PRL,Rossini_2019_PRA,Rossini_2007_PRA,Yuan_2007_PRA,YuanZi-Gang_2007_PRA,Damski_2011_PRA,Vicari_2018_PRA}.
Extensions to dissipative baths have further shown that dissipation
can even protect coherence \cite{Onizhuk_2024_PRL}.

Here, we instead study a non-Hermitian environment described
by a complex transverse-field Ising chain with periodic boundary
conditions, 
\begin{align}
H_{\mathrm{NI}} & =-\sum_{j=1}^{N}\left[JZ_{j}Z_{j+1}+(h_{x}X_{j}+ih_{y}Y_{j})\right],\label{eq:Ising_model}
\end{align}
where $X$, $Y$, and $Z$ denote the Pauli operators, $J$ is
the interaction strength, and $\boldsymbol{h}\equiv(h_{x},h_{y})$
characterizes the strength of the complex transverse field in
the $(x,y)$ plane, with $h_{x}$ and $h_{y}$ real. The imaginary
component of the transverse field can be effectively implemented
via measurement and post-selection protocols \cite{wang2023_AX}.
The Hamiltonian $H_{\mathrm{NI}}$ assumes a real spectrum for
$|h_{y}|<|h_{x}|$, but becomes non-diagonalizable at $h_{y}=\pm h_{x}$,
corresponding to EPs \cite{wang2023_AX}. These EPs exhibit distinct
non-Hermitian features, such as nontrivial topology and dimensional
reduction \cite{Bergholtz2021_RMP,Ding2022_NRP,Hu2022_PRR},
while also resembling Hermitian critical points through gap closing.

Decoherence generally arises from the coupling between the system
and its environment. Since the environment considered here is
non-Hermitian, we allow for both Hermitian and non-Hermitian
system--environment couplings. The total Hamiltonian reads 
\begin{equation}
H=H_{\mathrm{NI}}-|1\rangle\langle1|\otimes\sum_{j=1}^{N}\left(\delta_{x}X_{j}+i\delta_{y}Y_{j}\right),\label{eq:total_Hamiltonian}
\end{equation}
where $|0\rangle$ and $|1\rangle$ denote the ground and excited
states of the qubit, respectively. The parameters $\delta_{x}$
and $\delta_{y}$ are real and characterize the strengths of
the Hermitian and non-Hermitian couplings.

In contrast to the conventional scenario---where both the environment
and the system--environment coupling are Hermitian and enhanced
decoherence is typically observed near the quantum critical point
\cite{QuanHT2006_PRL,Rossini_2019_PRA,Rossini_2007_PRA,Yuan_2007_PRA,YuanZi-Gang_2007_PRA,Damski_2011_PRA,Vicari_2018_PRA}---the
interplay between Hermitian and non-Hermitian dynamics in our
model leads to qualitatively different behavior. As we demonstrate
below, near the exceptional transitions of the non-Hermitian
environment, decoherence can instead be strongly suppressed.

\emph{Suppression of decoherence at EPs.}---To analyze the decoherence
of the qubit, we consider an initial state in which the qubit
is prepared in the superposition $(|0\rangle+|1\rangle)/\sqrt{2}$,
while the environment is in its ground state $|G\rangle_{\mathrm{E}}$.
We quantify the coherence of the qubit using the $\ell_{1}$-norm
of coherence, defined as $C[\rho(t)]\equiv\sum_{i\neq j}|\rho_{ij}(t)|$
\cite{Baumgratz_2014_PRL}. Here, $\rho(t)$ denotes the reduced
density matrix of the qubit at time $t$, obtained by tracing
out the environmental degrees of freedom, $\rho(t)=\text{tr}_{\mathrm{E}}\left[|\psi(t)\rangle\langle\psi(t)|/\langle\psi(t)|\psi(t)\rangle\right]$,
where $|\psi(t)\rangle$ is the state of the total system.

We first consider the case in which only Hermitian system--environment
coupling is present, i.e., $\delta_{y}=0$. As shown in Fig.~\ref{fig: non-Hermitian_Spin_Chains}(b2),
when the non-Hermitian environment $H_{\mathrm{NI}}$ approaches
its exceptional transition ($h_{y}\rightarrow h_{x}$), the coherence
of the qubit decays more rapidly. This behavior resembles the
enhanced decoherence observed near quantum critical points in
conventional Hermitian environment \cite{QuanHT2006_PRL,Rossini_2019_PRA,Rossini_2007_PRA,Yuan_2007_PRA,YuanZi-Gang_2007_PRA,Damski_2011_PRA,Vicari_2018_PRA}.

In contrast, when both Hermitian and non-Hermitian system--environment
couplings are present, the coherence dynamics change qualitatively.
Fig.~\ref{fig: non-Hermitian_Spin_Chains}(b1) shows the time
evolution of the coherence for equal coupling strengths, $\delta_{x}=\delta_{y}$.
As the environment approaches its exceptional transition ($h_{y}\rightarrow h_{x}$),
decoherence is not enhanced but instead strongly suppressed.
This result highlights a key distinction from Hermitian environments,
where decoherence is typically amplified near critical points.
In non-Hermitian environments, exceptional transitions can instead
lead to coherence protection.

To examine whether this suppression is specific to the above
non-Hermitian Ising environment, we next consider another non-Hermitian
environment which is formed by a non-Hermitian spin-$1/2$ Heisenberg
chain described by 
\begin{equation}
H_{\mathrm{NH}}=H_{\mathrm{XXX}}-\sum_{j=1}^{N}\left(h_{x}X_{j}+ih_{y}Y_{j}\right),\label{eq:Heisenberg_model}
\end{equation}
where $H_{\mathrm{XXX}}\equiv-J\sum_{j=1}^{N}(X_{j}X_{j+1}+Y_{j}Y_{j+1}+Z_{j}Z_{j+1})$
is the isotropic Heisenberg Hamiltonian with periodic boundary
conditions. As in the Ising case, $H_{\mathrm{NH}}$ assumes
a completely real spectrum for $|h_{y}|<|h_{x}|$ and exhibits
EPs at $h_{y}=\pm h_{x}$ \cite{wang2023_AX}. We consider the
same form of system--environment coupling as in the non-Hermitian
Ising case {[}see the second term in Eq.~(\ref{eq:total_Hamiltonian}){]}.
The coherence dynamics for $\delta_{x}=\delta_{y}$ are shown
in Fig.~\ref{fig: non-Hermitian_Spin_Chains}(c1). Similar to
the Ising case, decoherence is strongly suppressed as the environment
approaches its exceptional transition. This suggests that decoherence
suppression near exceptional points is a generic feature of non-Hermitian
environments.

To understand the underlying mechanism, we note that in the absence
of the transverse field, the ferromagnetic Ising and Heisenberg
chains both possess a doubly degenerate ground-state manifold
consisting of fully polarized states. The complex transverse
field term, $-\sum_{j=1}^{N}(h_{x}X_{j}+ih_{y}Y_{j})$, lifts
this degeneracy. In particular, at the EPs (e.g., $h_{x}=h_{y}$),
a unique polarized ground state $|G\rangle_{\mathrm{E}}$ is
selected. In general, system--environment coupling can drive
the environment away from this polarized ground state, inducing
strong environmental fluctuations and thereby enhancing decoherence
near the exceptional transition. However, when the Hermitian
and the non-Hermitian system-environment couplings assume the
same strength, we find that the coupling cannot drive the environment
away from $|G\rangle_{\mathrm{E}}$ (see Supplementary Material~\cite{Sup_Mat}
for details). As a result, environmental fluctuations are suppressed,
leading to reduced decoherence. Away from the EPs, an analytic
treatment becomes intractable; nevertheless, extensive numerical
simulations confirm that decoherence suppression remains robust
in their vicinity (see Supplementary Material \cite{Sup_Mat}).

\emph{Suppression of decoherence at EPs of a non-Hermitian ultracold-gas
environment.}---The preceding section demonstrated that decoherence
suppression at EPs is a generic feature of non-Hermitian environments,
although the discussion was restricted to spin systems. We now
turn to a qualitatively different class of non-Hermitian environments
formed by ultracold gases in optical lattices, which are closely
connected to experimental realizations of non-Hermitian quantum
systems \cite{Jordens_2008_Nature,Mazurenko_2017_Nature,Boll_2016_Science,Schreiber_2015_Science}.
Specifically, we consider a non-Hermitian environment consisting
of ultracold Fermi gases in a one-dimensional optical lattice,
described by the Hamiltonian 
\begin{align}
H_{\mathrm{NF}}= & -J\sum_{j}^{N}\sum_{\sigma=\uparrow,\downarrow}(c_{j,\sigma}^{\dagger}c_{j+1,\sigma}+\mathrm{h.c.})+U\sum_{j=1}^{N}n_{j,\uparrow}n_{j,\downarrow}\nonumber \\
 & +h_{x}\sum_{j=1}^{N}(c_{j,\downarrow}^{\dagger}c_{j,\uparrow}+\mathrm{h.c.})-ih_{y}\sum_{j=1}^{N}(n_{j,\uparrow}-n_{j,\downarrow}),\label{eq:fermi_model}
\end{align}
where $c_{j,\sigma}^{\dagger}$ creates a fermion with spin $\sigma\in\{\uparrow,\downarrow\}$
at site $j$, and $n_{j,\sigma}\equiv c_{j,\sigma}^{\dagger}c_{j,\sigma}$.
The first two terms describe nearest-neighbor hopping with amplitude
$J$ and on-site interaction with strength $U$, respectively.
The third term represents spin-flip (Rabi) processes, while the
last term introduces spin-dependent gain and loss, rendering
the Hamiltonian non-Hermitian. The spectrum of $H_{\mathrm{NF}}$
is real for $|h_{y}|<|h_{x}|$, and an EP occurs at $h_{y}=h_{x}$,
where the Hamiltonian $H_{\mathrm{NF}}$ becomes non-diagonalizable
\cite{wang2023_AX}.

We investigate the coherence dynamics of a qubit coupled to this
non-Hermitian Fermi gas through the total Hamiltonian $H_{\mathrm{tot}}=H_{\mathrm{NF}}+|1\rangle\langle1|\otimes\sum_{j=1}^{N}\left[\delta_{x}(c_{j,\downarrow}^{\dagger}c_{j,\uparrow}+\mathrm{h.c.})-i\delta_{y}(n_{j,\uparrow}-n_{j,\downarrow})\right]$.
This form models a situation where the environment evolves under
$H_{\mathrm{NF}}$ when the qubit is in the ground state $|0\rangle$,
but experiences modified Rabi driving and gain--loss processes
when the qubit is in the excited state $|1\rangle$.

The resulting coherence dynamics are shown in Fig.~\ref{fig: non-Hermitian_fermi gases}.
We observe behavior qualitatively similar to that found in the
non-Hermitian spin-chain environments. In particular, when $\delta_{x}=\delta_{y}$,
decoherence is strongly suppressed at the exceptional transition
of the non-Hermitian Fermi gas environment {[}Fig.~\ref{fig: non-Hermitian_fermi gases}(a){]}.
These results identify ultracold Fermi gases as a promising experimental
platform for observing decoherence suppression at EPs.

\emph{Observability on digital quantum computers.}---In view
of the rapid progress in digital quantum computing, the predicted
coherence behavior should be experimentally accessible using
current quantum simulators. As a concrete demonstration, we simulate
the coherence dynamics of a system qubit coupled to a two-site
non-Hermitian Ising chain serving as its environment, implemented
as a quantum circuit in IBM Qiskit \cite{cross_2018_ibm}. For
the initial qubit state $\left(|0\rangle+|1\rangle\right)/\sqrt{2}$,
the coherence $C[\rho(t)]$ is equivalent to the overlap between
two time-evolved states of the environment (see \cite{Sup_Mat}).
This relation allows us to reduce the required quantum resources
by simulating only the environmental dynamics.

The quantum circuit implementing the time evolution of the two-site
non-Hermitian Ising chain is shown in Fig.~\ref{fig:IMBQ}(a).
The environmental ground state $|G\rangle_{\mathrm{E}}$ is prepared
by applying a two-qubit unitary operator $U_{G}$ to the initial
state $|0\rangle\otimes|0\rangle$. The operator $U_{G}$ can
be implemented by three elementary single-qubit gates combined
with a controlled-Z gate \cite{Sup_Mat}.

For a short time step $dt$, the time-evolution operator is Trotter
decomposed into a unitary part $U(dt)\equiv e^{idtJ\sum_{j=1}^{N}Z_{j}Z_{j+1}}e^{idt(h_{x}+\delta_{x})\sum_{j=1}^{N}X_{j}}$
and a non-unitary part $e^{-dt(h_{y}+\delta_{y})\sum_{j=1}^{N}Y_{j}}$.
The unitary part $U(dt)$ can be realized by standard one- and
two-qubit gates. The nonunitary part is realized using a measurement-based
protocol \cite{Brun_2008_PRA,wang_2025_PRB} that employs an
ancilla qubit {[}bottom row of Fig.~\ref{fig:IMBQ}(a){]} together
with projective measurement and postselection. Specifically,
a single qubit unitary operator $R\equiv(e^{i\pi/4}I+e^{-i\pi/4}\sum_{a=X,Y,Z}a)/2$
is applied to each qubit, followed by a CNOT gate and the inverse
operation $R^{-1}$. The ancilla qubit is initialized via the
rotation $U_{A}\equiv e^{-i(\pi/4+(h_{y}+\delta_{y})dt)Y}$.
After the CNOT interaction, the ancilla is measured in the computational
($Z$) basis and the outcome $|0\rangle$ is postselected, as
illustrated in Fig.~\ref{fig:IMBQ}(a). Finally, the overlap
between two time-evolved environmental states is obtained by
applying $U_{G}^{-1}$ and collecting the final measurement statistics
(see \cite{Sup_Mat} for more details).

Fig.~\ref{fig:IMBQ}(b) demonstrates quantum circuit simulations
of the coherence dynamics on IBM Qiskit \cite{cross_2018_ibm}
(discrete markers) and coherence dynamics by direct numerical
simulations (solid curves). When the Hermitian and non-Hermitian
system-environment couplings assume the same strength, the simulated
coherence dynamics clearly exhibit decoherence suppression as
the non-Hermitian environment approaches its exceptional transition,
in agreement with our theoretical predictions.

\emph{Conclusion}.---We have shown that decoherence induced
by non-Hermitian environments exhibits behavior fundamentally
distinct from that near Hermitian quantum critical points. Whereas
Hermitian criticality typically enhances decoherence, approaching
EPs of non-Hermitian environments can either enhance or strongly
suppress decoherence, depending on the balance between Hermitian
and non-Hermitian system--environment couplings. This establishes
non-Hermitian criticality as a qualitatively new regime for decoherence
dynamics. We demonstrated the generality of this effect across
diverse non-Hermitian environments, including spin chains and
interacting ultracold Fermi gases, and showed that the predicted
behavior is directly accessible on current quantum simulation
platforms. Our findings uncover an unexplored role of EPs in
open quantum dynamics and identify non-Hermitian environments
as a new avenue for controlling decoherence. More broadly, they
suggest that engineered non-Hermitian criticality may serve as
a practical resource for coherence protection in quantum information
processing, sensing, and quantum simulation. 
\begin{acknowledgments}
We acknowledge the use of IBM Quantum services for this work.
The views expressed are those of the authors, and do not reflect
the official policy or position of IBM or the IBM Quantum team.
This work is supported by the NKRDPC (Grant No.~2022YFA1405304),
NSFC (Grant No.~12275089), Guangdong Basic and Applied Basic
Research Foundation (Grants No.~2023A1515012800), Guangdong
Provincial Quantum Science Strategic Initiative (Grant No. GDZX2204003),
and Guangdong Provincial Key Laboratory (Grant No.~2020B1212060066). 
\end{acknowledgments}

\bibliography{ref}

@article{Zurek2003_RMP,
  title = {Decoherence, einselection, and the quantum origins of the classical},
  author = {Zurek, Wojciech Hubert},
  journal = {Rev. Mod. Phys.},
  volume = {75},
  issue = {3},
  pages = {715--775},
  numpages = {0},
  year = {2003},
  month = {May},
  publisher = {American Physical Society},
  doi = {10.1103/RevModPhys.75.715},
  url = {https://link.aps.org/doi/10.1103/RevModPhys.75.715}
}

@article{Schlosshauer_RMP_2005,
	title = {Decoherence, the measurement problem, and interpretations of quantum mechanics},
	author = {Schlosshauer, Maximilian},
	journal = {Rev. Mod. Phys.},
	volume = {76},
	issue = {4},
	pages = {1267--1305},
	numpages = {0},
	year = {2005},
	publisher = {American Physical Society},
	doi = {10.1103/RevModPhys.76.1267},
	url = {https://link.aps.org/doi/10.1103/RevModPhys.76.1267}
}

@book{Nielsen_Chuang_Book_2010, 
	place={Cambridge}, 
	title={Quantum Computation and Quantum Information: 10th Anniversary Edition}, publisher={Cambridge University Press}, 
	author={Nielsen, Michael A. and Chuang, Isaac L.}, year={2010}
	}

@article{Caldeira_PRL_1981,
	title = {Influence of Dissipation on Quantum Tunneling in Macroscopic Systems},
	author = {Caldeira, A. O. and Leggett, A. J.},
	journal = {Phys. Rev. Lett.},
	volume = {46},
	issue = {4},
	pages = {211--214},
	numpages = {0},
	year = {1981},
	month = {Jan},
	publisher = {American Physical Society},
	doi = {10.1103/PhysRevLett.46.211},
	url = {https://link.aps.org/doi/10.1103/PhysRevLett.46.211}
}

@article{Leggett_RMP_1987,
		title = {Dynamics of the dissipative two-state system},
		author = {Leggett, A. J. and Chakravarty, S. and Dorsey, A. T. and Fisher, Matthew P. A. and Garg, Anupam and Zwerger, W.},
		journal = {Rev. Mod. Phys.},
		volume = {59},
		issue = {1},
		pages = {1--85},
		numpages = {0},
		year = {1987},
		month = {Jan},
		publisher = {American Physical Society},
		doi = {10.1103/RevModPhys.59.1},
		url = {https://link.aps.org/doi/10.1103/RevModPhys.59.1}
}

@article{Harrington_Nat_Rev_Phys_2022,
	author = {Harrington, Patrick M. and Mueller, Erich J. and Murch, Kater W.},
	date = {2022/10/01},
	doi = {10.1038/s42254-022-00494-8},
	id = {Harrington2022},
	isbn = {2522-5820},
	journal = {Nature Reviews Physics},
	number = {10},
	pages = {660--671},
	title = {Engineered dissipation for quantum information science},
	url = {https://doi.org/10.1038/s42254-022-00494-8},
	volume = {4},
	year = {2022}
	}

@article{Fruchart_Nature_2021,
		author = {Fruchart, Michel and Hanai, Ryo and Littlewood, Peter B. and Vitelli, Vincenzo},
		date = {2021/04/01},
		doi = {10.1038/s41586-021-03375-9},
		id = {Fruchart2021},
		isbn = {1476-4687},
		journal = {Nature},
		number = {7854},
		pages = {363--369},
		title = {Non-reciprocal phase transitions},
		url = {https://doi.org/10.1038/s41586-021-03375-9},
		volume = {592},
		year = {2021}
}

@article{QuanHT2006_PRL,
  title = {Decay of Loschmidt Echo Enhanced by Quantum Criticality},
  author = {Quan, H. T. and Song, Z. and Liu, X. F. and Zanardi, P. and Sun, C. P.},
  journal = {Phys. Rev. Lett.},
  volume = {96},
  issue = {14},
  pages = {140604},
  numpages = {4},
  year = {2006},
  month = {Apr},
  publisher = {American Physical Society},
  doi = {10.1103/PhysRevLett.96.140604},
  url = {https://link.aps.org/doi/10.1103/PhysRevLett.96.140604}
}

@article{Zhang2008_PRL,
  title = {Detection of Quantum Critical Points by a Probe Qubit},
  author = {Zhang, Jingfu and Peng, Xinhua and Rajendran, Nageswaran and Suter, Dieter},
  journal = {Phys. Rev. Lett.},
  volume = {100},
  issue = {10},
  pages = {100501},
  numpages = {4},
  year = {2008},
  month = {Mar},
  publisher = {American Physical Society},
  doi = {10.1103/PhysRevLett.100.100501},
  url = {https://link.aps.org/doi/10.1103/PhysRevLett.100.100501}
}

@article{ashida2020_AP,
author = {Yuto Ashida, Zongping Gong and Masahito Ueda},
title = {Non-{H}ermitian physics},
journal = {Adv. Phys.},
volume = {69},
number = {3},
pages = {249--435},
year = {2020},
publisher = {Taylor \& Francis},
doi = {10.1080/00018732.2021.1876991},
URL = {https://doi.org/10.1080/00018732.2021.1876991}
}

@misc{wang2023_AX,
  title={Measurement-induced integer families of critical dynamical scaling in quantum many-body systems}, 
  author={Zuo Wang and Shi-Liang Zhu and Li-Jun Lang and Liang He},
  howpublished = {\href{https://arxiv.org/abs/2308.06567}{arXiv:2308.06567}},
  }

@article{Bergholtz2021_RMP,
  title = {Exceptional topology of non-Hermitian systems},
  author = {Bergholtz, Emil J. and Budich, Jan Carl and Kunst, Flore K.},
  journal = {Rev. Mod. Phys.},
  volume = {93},
  issue = {1},
  pages = {015005},
  numpages = {31},
  year = {2021},
  month = {Feb},
  publisher = {American Physical Society},
  doi = {10.1103/RevModPhys.93.015005},
  url = {https://link.aps.org/doi/10.1103/RevModPhys.93.015005}
}

@article{Ding2022_NRP,
  author    = {Ding, Kun and Fang, Chen and Ma, Guancong},
  title     = {Non-Hermitian topology and exceptional-point geometries},
  journal   = {Nat. Rev. Phys.},
  volume    = {4},
  number    = {12},
  pages     = {745--760},
  year      = {2022},
  doi       = {10.1038/s42254-022-00516-5},
  url       = {https://doi.org/10.1038/s42254-022-00516-5},
  }

@article{Hu2022_PRR,
  title = {Knot topology of exceptional point and non-Hermitian no-go theorem},
  author = {Hu, Haiping and Sun, Shikang and Chen, Shu},
  journal = {Phys. Rev. Res.},
  volume = {4},
  issue = {2},
  pages = {L022064},
  numpages = {6},
  year = {2022},
  month = {Jun},
  publisher = {American Physical Society},
  doi = {10.1103/PhysRevResearch.4.L022064},
  url = {https://link.aps.org/doi/10.1103/PhysRevResearch.4.L022064}
}

@article{wang_2025_PRB,
	title = {Emergent non-Hermitian conservation laws at exceptional points},
	author = {Wang, Zuo and He, Liang},
	journal = {Phys. Rev. B},
	volume = {111},
	issue = {10},
	pages = {L100305},
	numpages = {6},
	year = {2025},
	month = {Mar},
	publisher = {American Physical Society},
	doi = {10.1103/PhysRevB.111.L100305},
	url = {https://link.aps.org/doi/10.1103/PhysRevB.111.L100305}
}

@article{Sup_Mat,
	Journal = {See Supplemental Material for a discussion on relevant technical details}
}

@article{Jordens_2008_Nature,
  author    = {Jördens, Robert and Strohmaier, Niels and Günter, Kenneth and Moritz, Henning and Esslinger, Tilman},
  title     = {A Mott insulator of fermionic atoms in an optical lattice},
  journal   = {Nature},
  year      = {2008},
  volume    = {455},
  number    = {7210},
  pages     = {204--207},
  doi       = {10.1038/nature07244},
  url       = {https://doi.org/10.1038/nature07244},
  issn      = {1476-4687}
  }

@article{Schreiber_2015_Science,
  author    = {Michael Schreiber and Sean S. Hodgman and Pranjal Bordia and Henrik P. Lüschen and Mark H. Fischer and Ronen Vosk and Ehud Altman and Ulrich Schneider and Immanuel Bloch},
  title     = {Observation of many-body localization of interacting fermions in a quasirandom optical lattice},
  journal   = {Science},
  volume    = {349},
  number    = {6250},
  pages     = {842--845},
  year      = {2015},
  doi       = {10.1126/science.aaa7432},
  url       = {https://www.science.org/doi/abs/10.1126/science.aaa7432}
}

@article{Boll_2016_Science,
  author    = {Martin Boll and Timon A. Hilker and Guillaume Salomon and Ahmed Omran and Jacopo Nespolo and Lode Pollet and Immanuel Bloch and Christian Gross},
  title     = {Spin- and density-resolved microscopy of antiferromagnetic correlations in Fermi-Hubbard chains},
  journal   = {Science},
  volume    = {353},
  number    = {6305},
  pages     = {1257--1260},
  year      = {2016},
  doi       = {10.1126/science.aag1635},
  url       = {https://www.science.org/doi/abs/10.1126/science.aag1635}
}

@article{Mazurenko_2017_Nature,
  author    = {Anton Mazurenko and Christie S. Chiu and Geoffrey Ji and Maxwell F. Parsons and Márton Kanász-Nagy and Richard Schmidt and Fabian Grusdt and Eugene Demler and Daniel Greif and Markus Greiner},
  title     = {A cold-atom Fermi–Hubbard antiferromagnet},
  journal   = {Nature},
  volume    = {545},
  number    = {7655},
  pages     = {462--466},
  year      = {2017},
  doi       = {10.1038/nature22362},
  url       = {https://doi.org/10.1038/nature22362},
  issn      = {1476-4687}
}

@article{Brun_2008_PRA,
  title = {Test of weak measurement on a two- or three-qubit computer},
  author = {Brun, Todd A. and Di\'osi, Lajos and Strunz, Walter T.},
  journal = {Phys. Rev. A},
  volume = {77},
  issue = {3},
  pages = {032101},
  numpages = {6},
  year = {2008},
  month = {Mar},
  publisher = {American Physical Society},
  doi = {10.1103/PhysRevA.77.032101},
  url = {https://link.aps.org/doi/10.1103/PhysRevA.77.032101}
}

@inproceedings{cross_2018_ibm,
  title={The IBM Q experience and QISKit open-source quantum computing software},
  author={Cross, Andrew},
  booktitle={APS March meeting abstracts},
  volume={2018},
  pages={L58--003},
  year={2018}
}

@article{Zhao_2023_PRX,
  title = {Making Trotterization Adaptive and Energy-Self-Correcting for NISQ Devices and Beyond},
  author = {Zhao, Hongzheng and Bukov, Marin and Heyl, Markus and Moessner, Roderich},
  journal = {PRX Quantum},
  volume = {4},
  issue = {3},
  pages = {030319},
  numpages = {11},
  year = {2023},
  month = {Aug},
  publisher = {American Physical Society},
  doi = {10.1103/PRXQuantum.4.030319},
  url = {https://link.aps.org/doi/10.1103/PRXQuantum.4.030319}
}

@article{Zhao_2024_PRL,
  title = {Adaptive Trotterization for Time-Dependent Hamiltonian Quantum Dynamics Using Piecewise Conservation Laws},
  author = {Zhao, Hongzheng and Bukov, Marin and Heyl, Markus and Moessner, Roderich},
  journal = {Phys. Rev. Lett.},
  volume = {133},
  issue = {1},
  pages = {010603},
  numpages = {8},
  year = {2024},
  month = {Jul},
  publisher = {American Physical Society},
  doi = {10.1103/PhysRevLett.133.010603},
  url = {https://link.aps.org/doi/10.1103/PhysRevLett.133.010603}
}

@article{Davide_2021_PR,
title = {Coherent and dissipative dynamics at quantum phase transitions},
journal = {Phys. Rep.},
volume = {936},
pages = {1-110},
year = {2021},
note = {Coherent and dissipative dynamics at quantum phase transitions},
issn = {0370-1573},
doi = {https://doi.org/10.1016/j.physrep.2021.08.003},
url = {https://www.sciencedirect.com/science/article/pii/S0370157321003380},
author = {Davide Rossini and Ettore Vicari},

}

@article{Rossini_2007_PRA,
  title = {Decoherence induced by interacting quantum spin baths},
  author = {Rossini, Davide and Calarco, Tommaso and Giovannetti, Vittorio and Montangero, Simone and Fazio, Rosario},
  journal = {Phys. Rev. A},
  volume = {75},
  issue = {3},
  pages = {032333},
  numpages = {16},
  year = {2007},
  month = {Mar},
  publisher = {American Physical Society},
  doi = {10.1103/PhysRevA.75.032333},
  url = {https://link.aps.org/doi/10.1103/PhysRevA.75.032333}
}

@article{Yuan_2007_PRA,
  title = {Loschmidt echo and Berry phase of a quantum system coupled to an $XY$ spin chain: Proximity to a quantum phase transition},
  author = {Yuan, Zi-Gang and Zhang, Ping and Li, Shu-Shen},
  journal = {Phys. Rev. A},
  volume = {75},
  issue = {1},
  pages = {012102},
  numpages = {7},
  year = {2007},
  month = {Jan},
  publisher = {American Physical Society},
  doi = {10.1103/PhysRevA.75.012102},
  url = {https://link.aps.org/doi/10.1103/PhysRevA.75.012102}
}

@article{YuanZi-Gang_2007_PRA,
  title = {Disentanglement of two qubits coupled to an $XY$ spin chain: Role of quantum phase transition},
  author = {Yuan, Zi-Gang and Zhang, Ping and Li, Shu-Shen},
  journal = {Phys. Rev. A},
  volume = {76},
  issue = {4},
  pages = {042118},
  numpages = {7},
  year = {2007},
  month = {Oct},
  publisher = {American Physical Society},
  doi = {10.1103/PhysRevA.76.042118},
  url = {https://link.aps.org/doi/10.1103/PhysRevA.76.042118}
}

@article{Damski_2011_PRA,
  title = {Critical dynamics of decoherence},
  author = {Damski, Bogdan and Quan, H. T. and Zurek, Wojciech H.},
  journal = {Phys. Rev. A},
  volume = {83},
  issue = {6},
  pages = {062104},
  numpages = {8},
  year = {2011},
  month = {Jun},
  publisher = {American Physical Society},
  doi = {10.1103/PhysRevA.83.062104},
  url = {https://link.aps.org/doi/10.1103/PhysRevA.83.062104}
}

@article{Vicari_2018_PRA,
  title = {Decoherence dynamics of qubits coupled to systems at quantum transitions},
  author = {Vicari, Ettore},
  journal = {Phys. Rev. A},
  volume = {98},
  issue = {5},
  pages = {052127},
  numpages = {8},
  year = {2018},
  month = {Nov},
  publisher = {American Physical Society},
  doi = {10.1103/PhysRevA.98.052127},
  url = {https://link.aps.org/doi/10.1103/PhysRevA.98.052127}
}

@article{Rossini_2019_PRA,
  title = {Scaling of decoherence and energy flow in interacting quantum spin systems},
  author = {Rossini, Davide and Vicari, Ettore},
  journal = {Phys. Rev. A},
  volume = {99},
  issue = {5},
  pages = {052113},
  numpages = {15},
  year = {2019},
  month = {May},
  publisher = {American Physical Society},
  doi = {10.1103/PhysRevA.99.052113},
  url = {https://link.aps.org/doi/10.1103/PhysRevA.99.052113}
}

@article{Onizhuk_2024_PRL,
  title = {Understanding Central Spin Decoherence Due to Interacting Dissipative Spin Baths},
  author = {Onizhuk, Mykyta and Wang, Yu-Xin and Nagura, Jonah and Clerk, Aashish A. and Galli, Giulia},
  journal = {Phys. Rev. Lett.},
  volume = {132},
  issue = {25},
  pages = {250401},
  numpages = {7},
  year = {2024},
  month = {Jun},
  publisher = {American Physical Society},
  doi = {10.1103/PhysRevLett.132.250401},
  url = {https://link.aps.org/doi/10.1103/PhysRevLett.132.250401}
}

@article{Baumgratz_2014_PRL,
  title = {Quantifying Coherence},
  author = {Baumgratz, T. and Cramer, M. and Plenio, M. B.},
  journal = {Phys. Rev. Lett.},
  volume = {113},
  issue = {14},
  pages = {140401},
  numpages = {5},
  year = {2014},
  month = {Sep},
  publisher = {American Physical Society},
  doi = {10.1103/PhysRevLett.113.140401},
  url = {https://link.aps.org/doi/10.1103/PhysRevLett.113.140401}
}

@article{G_1991_JPA,
doi = {10.1088/0305-4470/24/22/021},
url = {https://dx.doi.org/10.1088/0305-4470/24/22/021},
year = {1991},
month = {nov},
publisher = {},
volume = {24},
number = {22},
pages = {5371},
author = {G von Gehlen},
title = {Critical and off-critical conformal analysis of the Ising quantum chain in an imaginary field},
journal = {J. Phys. A: Math. Gen.},

}

@article{Uzelac_1979_PRL,
  title = {Yang-Lee Edge Singularity from a Real-Space Renormalization-Group Method},
  author = {Uzelac, K. and Pfeuty, P. and Jullien, R.},
  journal = {Phys. Rev. Lett.},
  volume = {43},
  issue = {12},
  pages = {805--808},
  numpages = {0},
  year = {1979},
  month = {Sep},
  publisher = {American Physical Society},
  doi = {10.1103/PhysRevLett.43.805},
  url = {https://link.aps.org/doi/10.1103/PhysRevLett.43.805}
}

@article{Castro_2009_JPA,
doi = {10.1088/1751-8113/42/46/465211},
url = {https://dx.doi.org/10.1088/1751-8113/42/46/465211},
year = {2009},
month = {oct},
publisher = {},
volume = {42},
number = {46},
pages = {465211},
author = {Castro-Alvaredo, Olalla A and Fring, Andreas},
title = {A spin chain model with non-Hermitian interaction: the Ising quantum spin chain in an imaginary field},
journal = {J. Phys. A: Math. Theor.}
}

@book{Breuer_2007_OUP,
    author = {Breuer, Heinz-Peter and Petruccione, Francesco},
    title = {The Theory of Open Quantum Systems},
    publisher = {Oxford University Press},
    year = {2007},
    month = {01},
    isbn = {9780199213900},
    doi = {10.1093/acprof:oso/9780199213900.001.0001},
    url = {https://doi.org/10.1093/acprof:oso/9780199213900.001.0001},
}

@article{Breuer_2016_RMP,
  title = {Colloquium: Non-Markovian dynamics in open quantum systems},
  author = {Breuer, Heinz-Peter and Laine, Elsi-Mari and Piilo, Jyrki and Vacchini, Bassano},
  journal = {Rev. Mod. Phys.},
  volume = {88},
  issue = {2},
  pages = {021002},
  numpages = {24},
  year = {2016},
  month = {Apr},
  publisher = {American Physical Society},
  doi = {10.1103/RevModPhys.88.021002},
  url = {https://link.aps.org/doi/10.1103/RevModPhys.88.021002}
}


\clearpage\onecolumngrid 

\vspace{\columnsep} 
\begin{center}
{\large\textbf{Supplemental Material for ``Suppression of Decoherence
at Exceptional Transitions''}}{\large\par}
\par\end{center}

\begin{center}
 
\par\end{center}

\vspace{\columnsep}


\twocolumngrid

\setcounter{equation}{0}

\setcounter{figure}{0}

\setcounter{page}{1}

\setcounter{section}{0}

\global\long\def\theequation{S\arabic{equation}}%
\global\long\def\thesection{S\arabic{section}}%
\global\long\def\thefigure{S\arabic{figure}}%
%
%
%
%
%
%
%

The supplemental material provides additional numerical and analytical
results supporting the main text. We present the coherence dynamics
of the qubit for relevant non-Hermitian system--environment
couplings and show that suppression of decoherence near the exceptional
transition is robust over a finite parameter region. We further
analyze the overlap of time-evolved environmental states and
demonstrate that the non-Hermitian dynamics dynamically selects
a unique polarized environmental ground state, thereby suppressing
decoherence. Finally, we provide implementation details for the
digital quantum simulation in IBM Qiskit, including state preparation,
ancilla-assisted postselection for nonunitary evolution, time-step
optimization, and data-processing procedures. 

\section{Decoherence Dynamics for Different Couplings}

\begin{figure}
\includegraphics[width=1.7in]{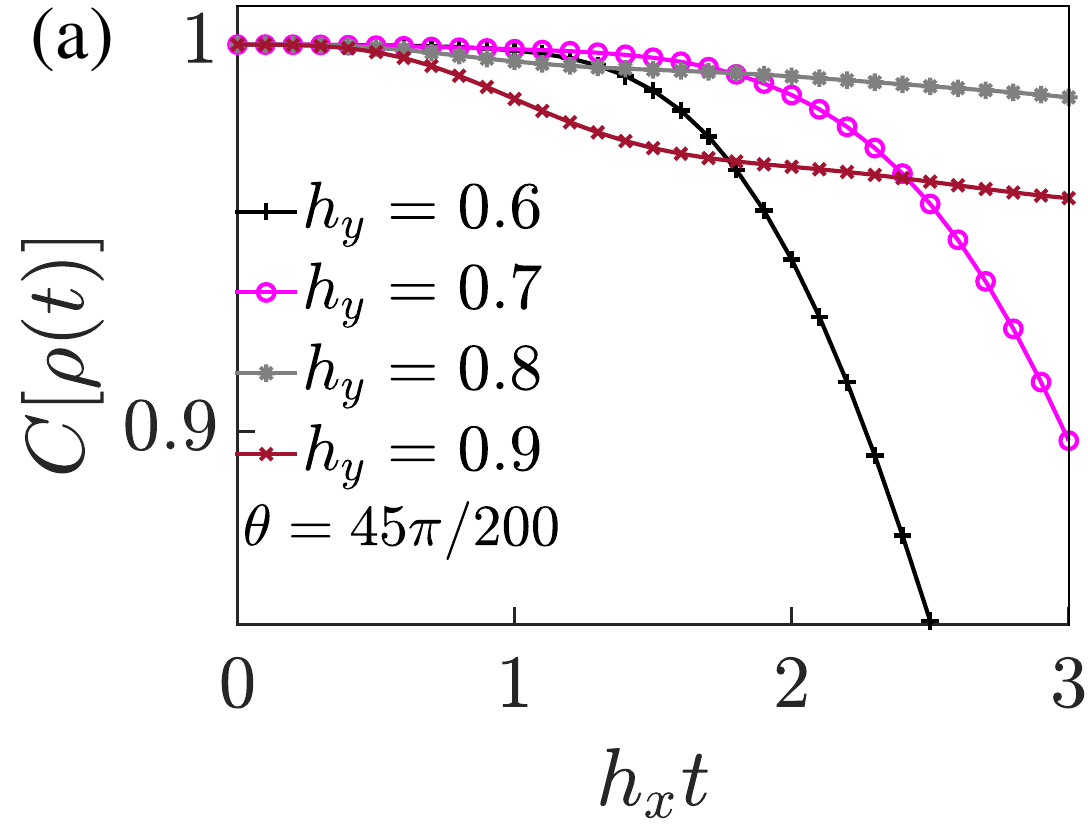}\includegraphics[width=1.7in]{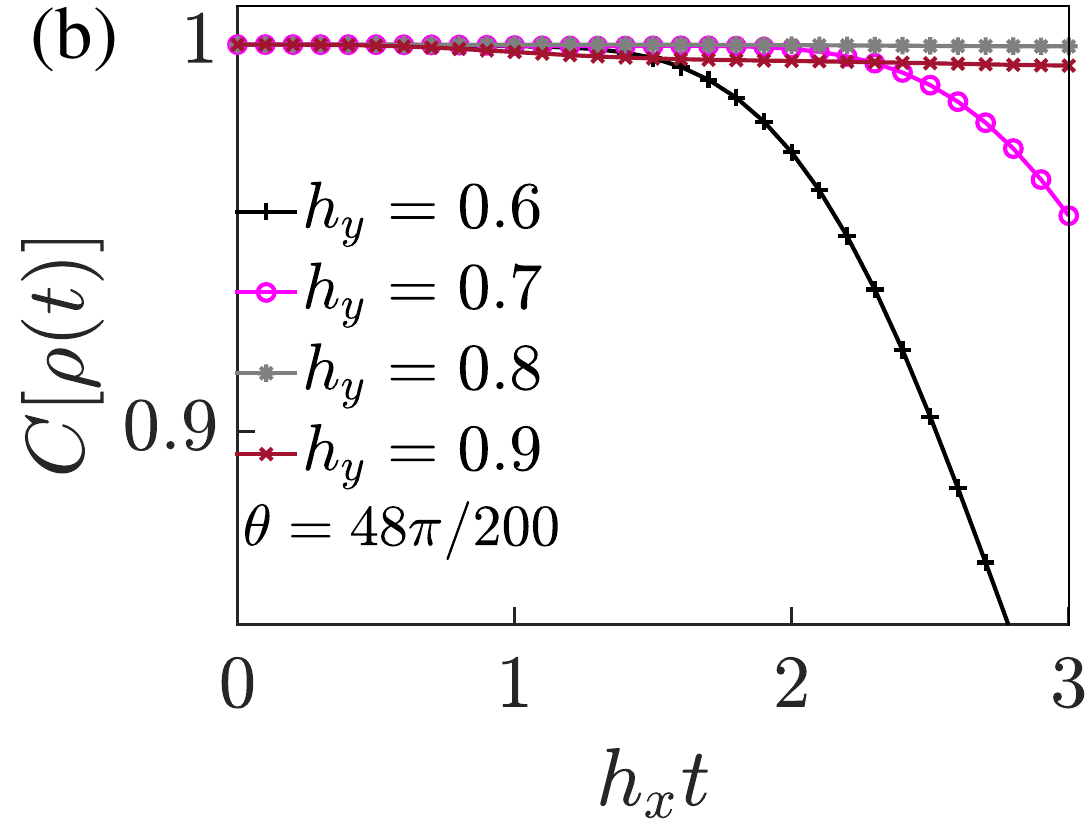}

\includegraphics[width=1.7in]{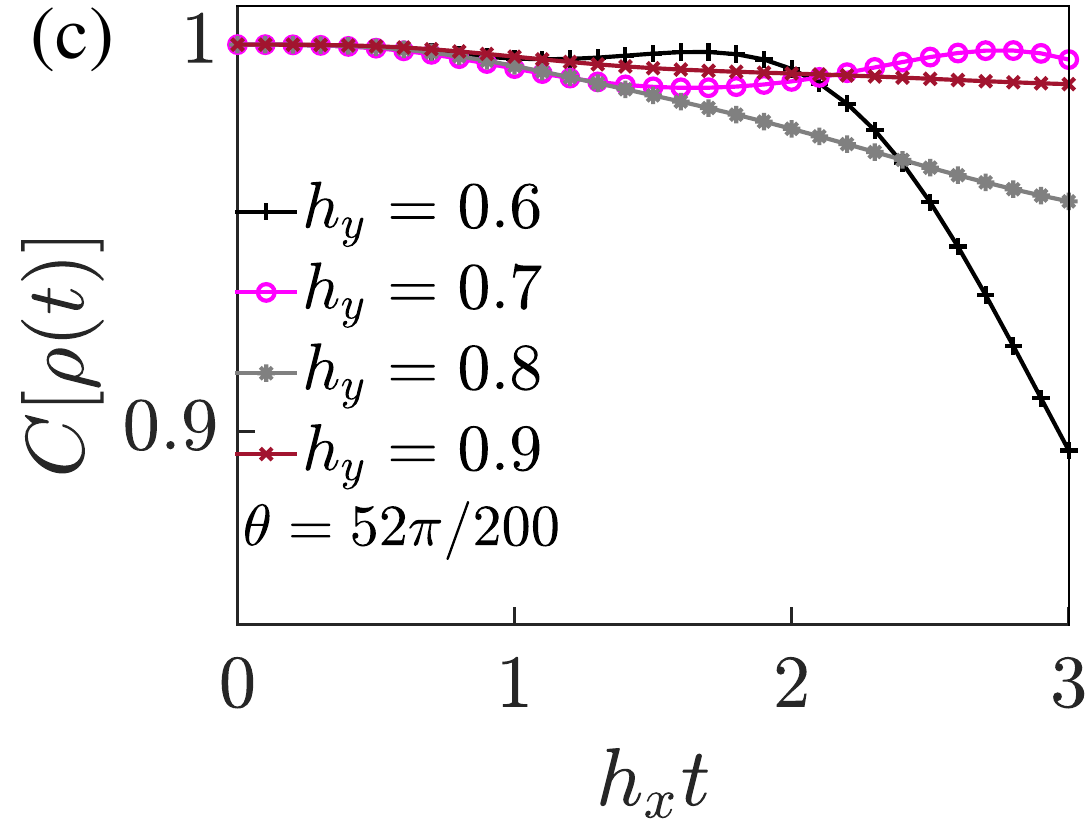}\includegraphics[width=1.7in]{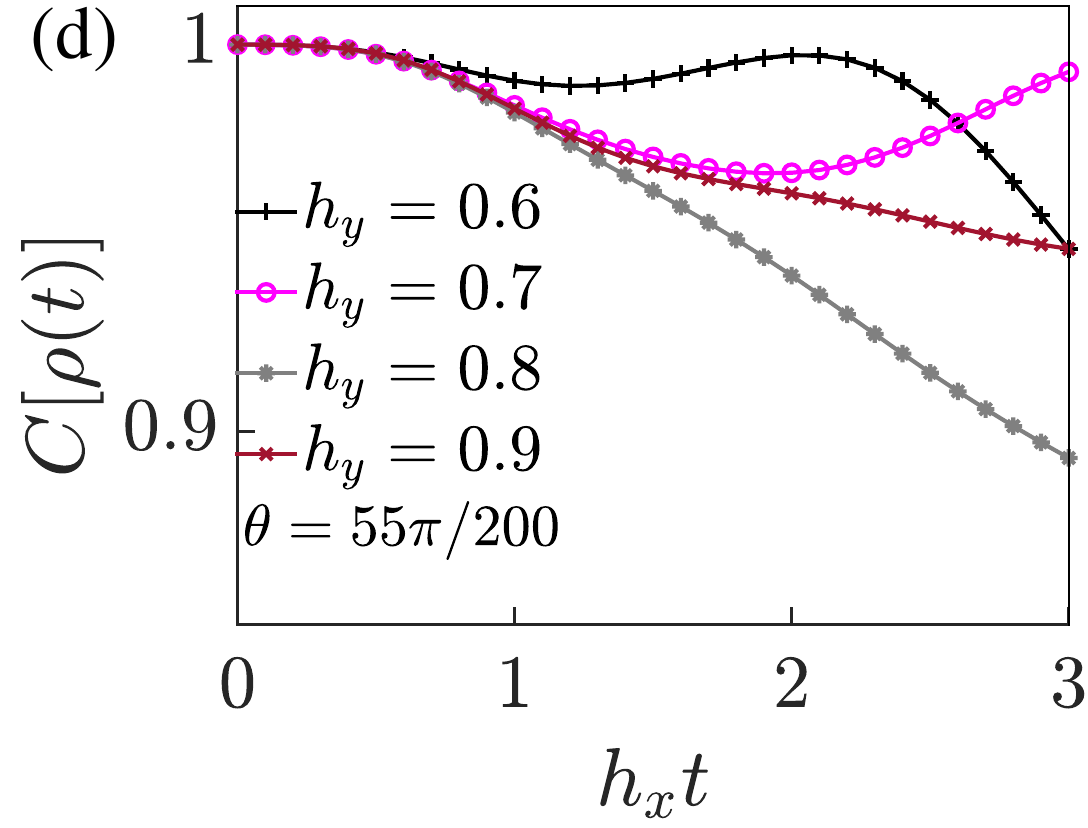}

\caption{Dynamics of the qubit coherence $C[\rho(t)]$ as a non-Hermitian
environment formed by a non-Hermitian Ising model under different
coupling strengths $\delta_{x}=|\boldsymbol{\delta}|\sin\theta$
and $\delta_{y}=|\boldsymbol{\delta}|\cos\theta$ approaches
its EPs, $h_{y}\rightarrow h_{x}=1$. (a-d) The $C[\rho(t)]$
for time $t\in[0,3]$ with $\theta=45\pi/200$ (a), $\theta=48\pi/200$
(b), $\theta=52\pi/200$ (c), $\theta=55\pi/200$ (d). All simulations
use $N=25$ and $J=1/2$, $h_{x}=1$, $|\boldsymbol{\delta}|=\sqrt{\delta_{x}^{2}+\delta_{y}^{2}}=0.1$.
See text for more details.}

\label{fig:transition_around_4_pi}
\end{figure}

\begin{figure}
\includegraphics[width=1.9in]{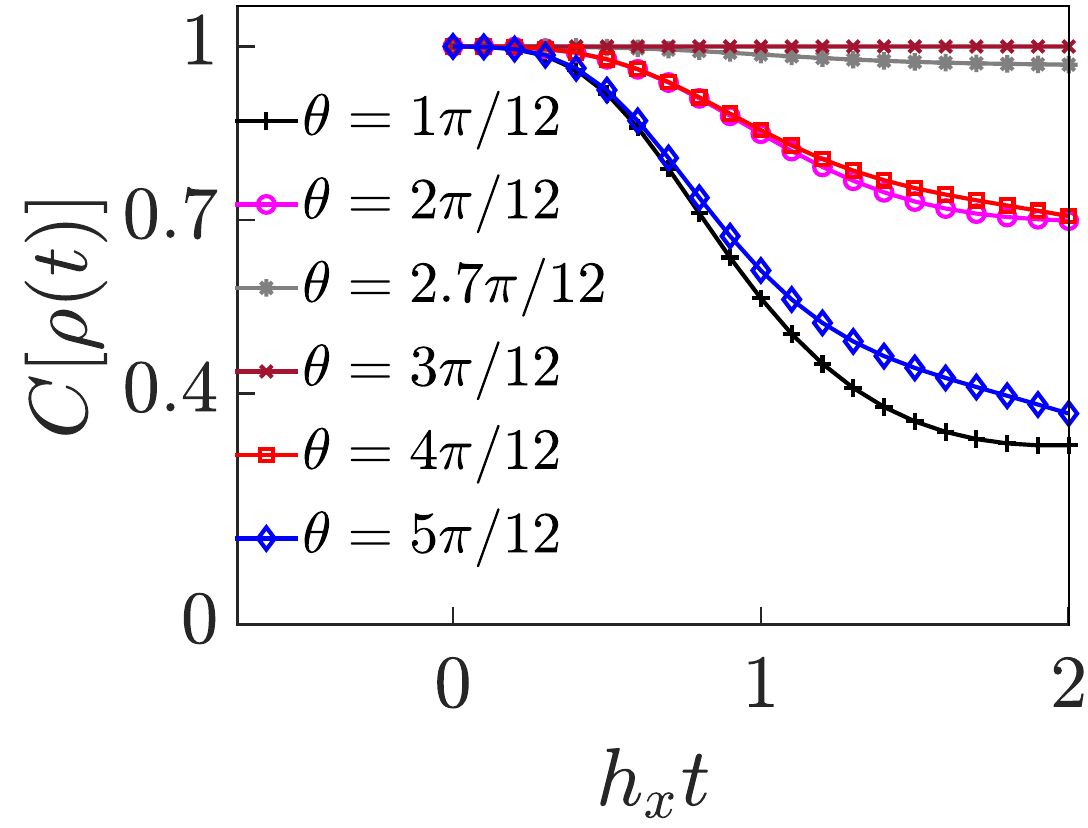}

\caption{Dynamics of the qubit coherence $C[\rho(t)]$ as a non-Hermitian
environment formed by a non-Hermitian Ising model with fixed
complex field $h_{y}$ under different $\theta$. All simulations
use $N=25$ and $h_{y}=0.9$, $J=1/2$, $h_{x}=1$ under different
couplings $\delta_{x}=|\boldsymbol{\delta}|\sin\theta$, $\delta_{y}=|\boldsymbol{\delta}|\cos\theta$,
$|\boldsymbol{\delta}|=\sqrt{\delta_{x}^{2}+\delta_{y}^{2}}=0.1$.
See text for more details.}

\label{fig:decay_fix_hy}
\end{figure}

In this section, we show that suppression of decoherence near
the exceptional transition at $\delta_{x}=\delta_{y}$ is robust
over a finite parameter region. We parameterize the non-Hermitian
system-environment couplings $\boldsymbol{\delta}$ by an angle
$\theta$ via $(\delta_{x},\delta_{y})=|\boldsymbol{\delta}|(\sin\theta,\cos\theta)$
and present the numerical simulation results that reveal the
detailed behavior of $C[\rho(t)]$ decay near EP, located around
$\theta=\pi/4$ in the real energy spectrum. Fig.~\ref{fig:transition_around_4_pi}
presents the dynamics of $C[\rho(t)]$ near EP, with particular
attention to the transitional regions between enhanced and suppressed
decay, which emerge as the angle is tuned across $\pi/4$. To
isolate the role of the complex field component $h_{y}$, we
fix its value at $h_{y}=0.9$, allowing us to focus on $C[\rho(t)]$
behavior within the range $\theta\in[0,\pi/2)$. As shown in
Fig.~\ref{fig:decay_fix_hy}, the $C[\rho(t)]$ decay exhibits
a symmetric distribution around $\theta=\pi/4$ near EP. Deviations
from either side of this angle lead to more enhanced decay, indicating
that $\theta=\pi/4$ corresponds to a special, decay-minimizing
coupling scenario. To demonstrate the robustness of this suppressed
decay of $C[\rho(t)]$, we also consider couplings with parameter
$\theta$ close to $\pi/4$, as shown in Fig.~\ref{fig:decay_fix_hy}.
These results confirm that the contrasting behavior exactly at
$\theta=\pi/4$ is not a fine-tuned artifact, but a physically
robust phenomenon observable in a finite parameter region.

\section{Ground-State Susceptibility Near the EPs}

\begin{figure}
\includegraphics[width=1.75in]{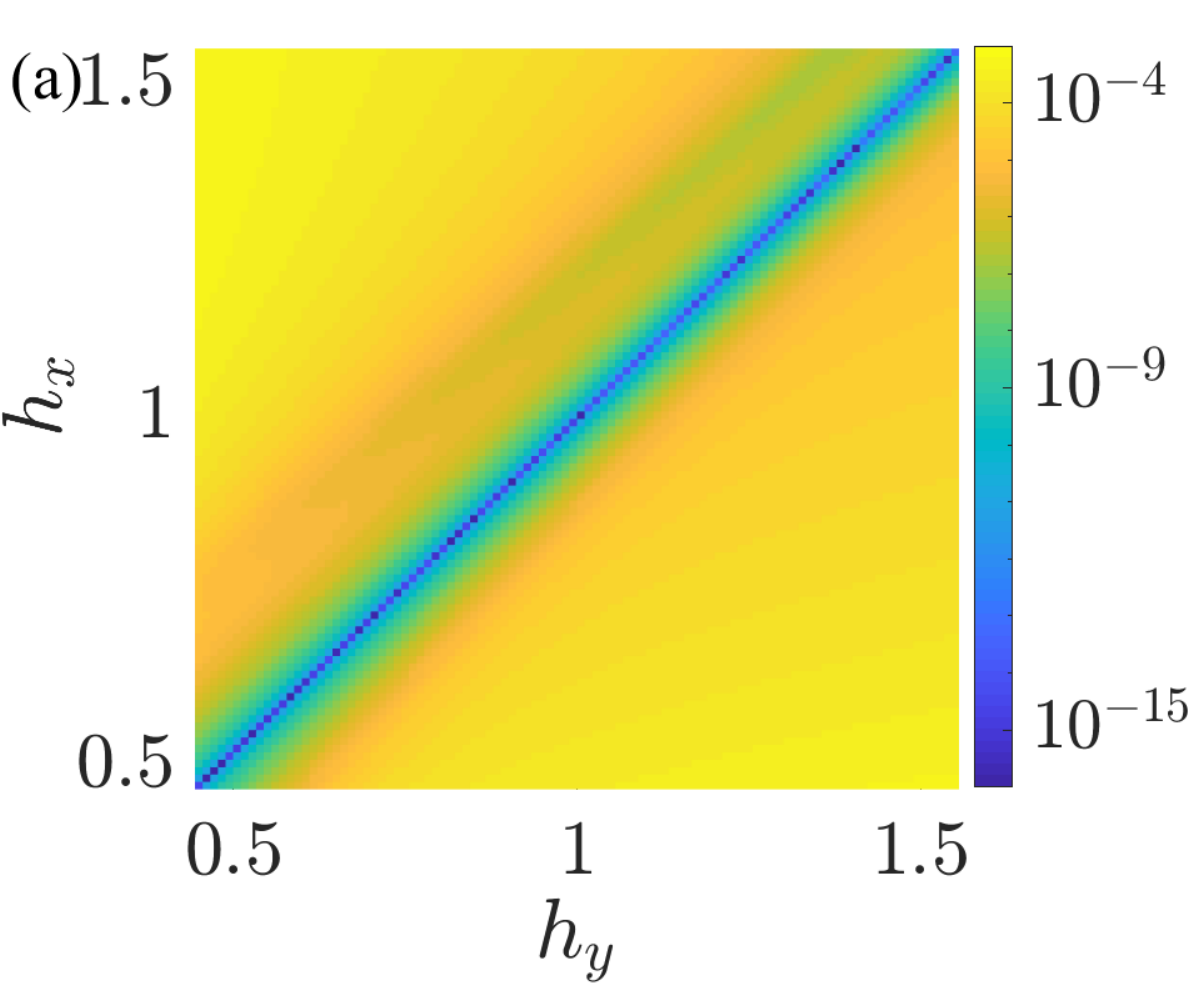}\includegraphics[width=1.75in]{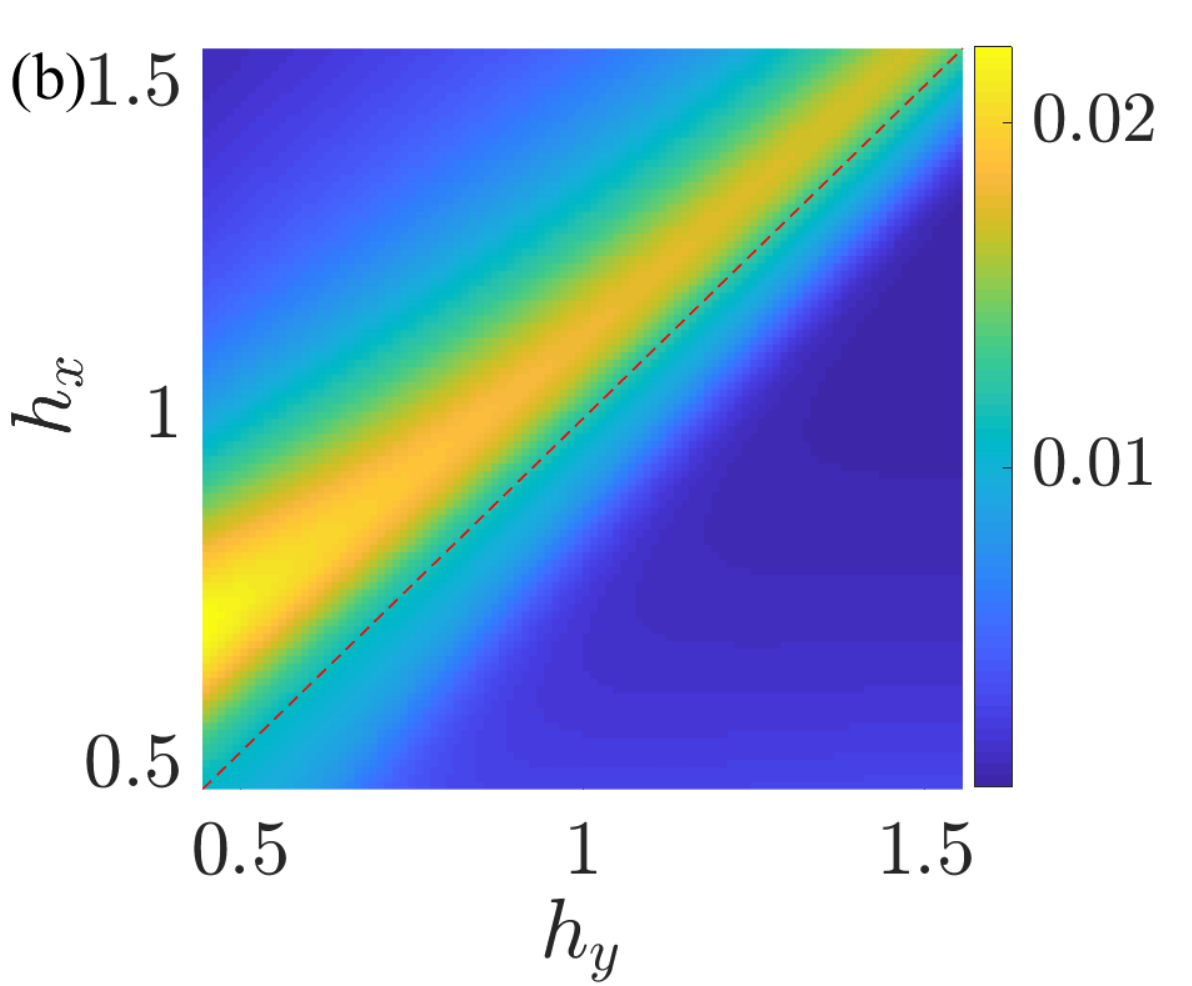}

\includegraphics[width=1.75in]{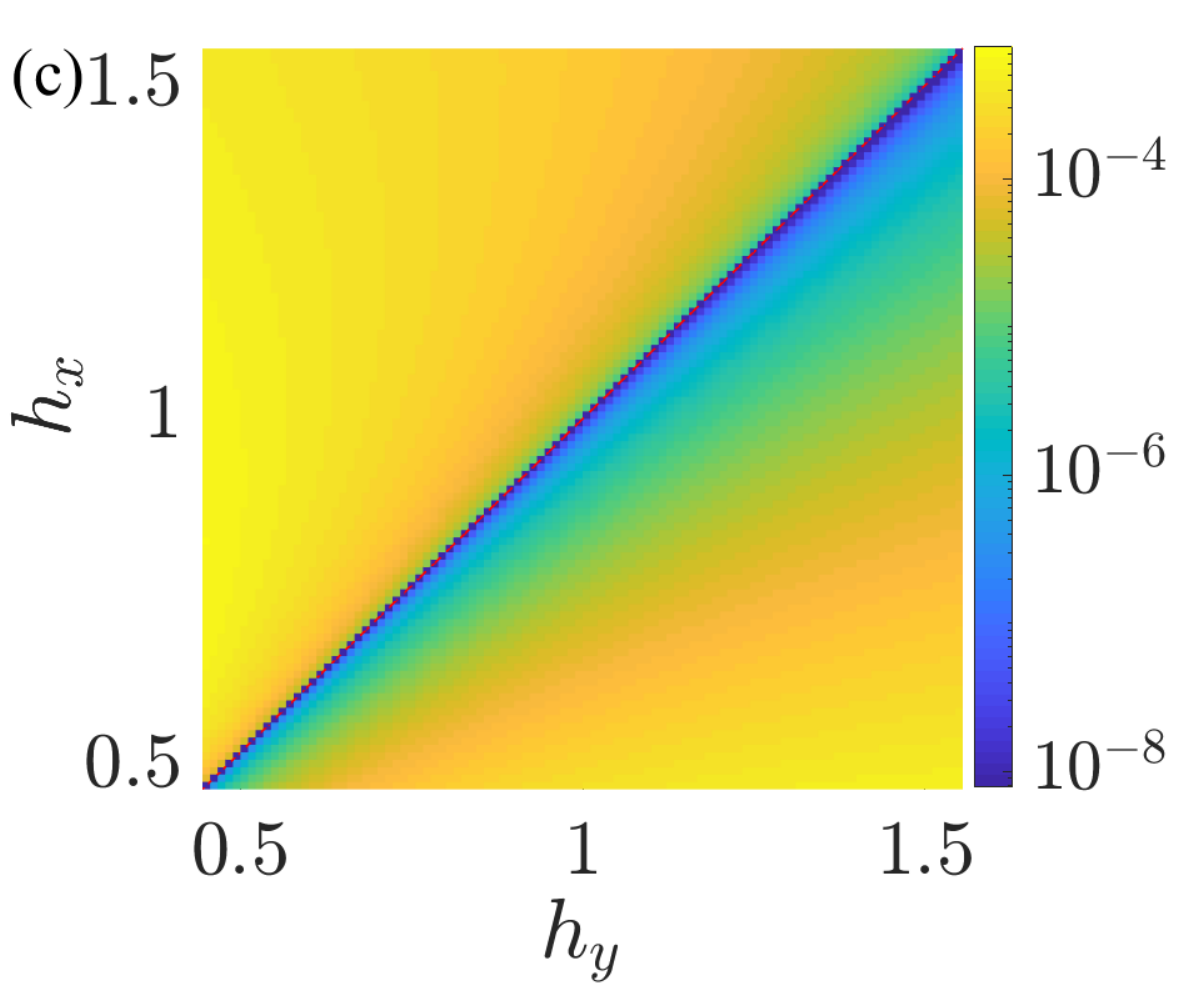}\includegraphics[width=1.75in]{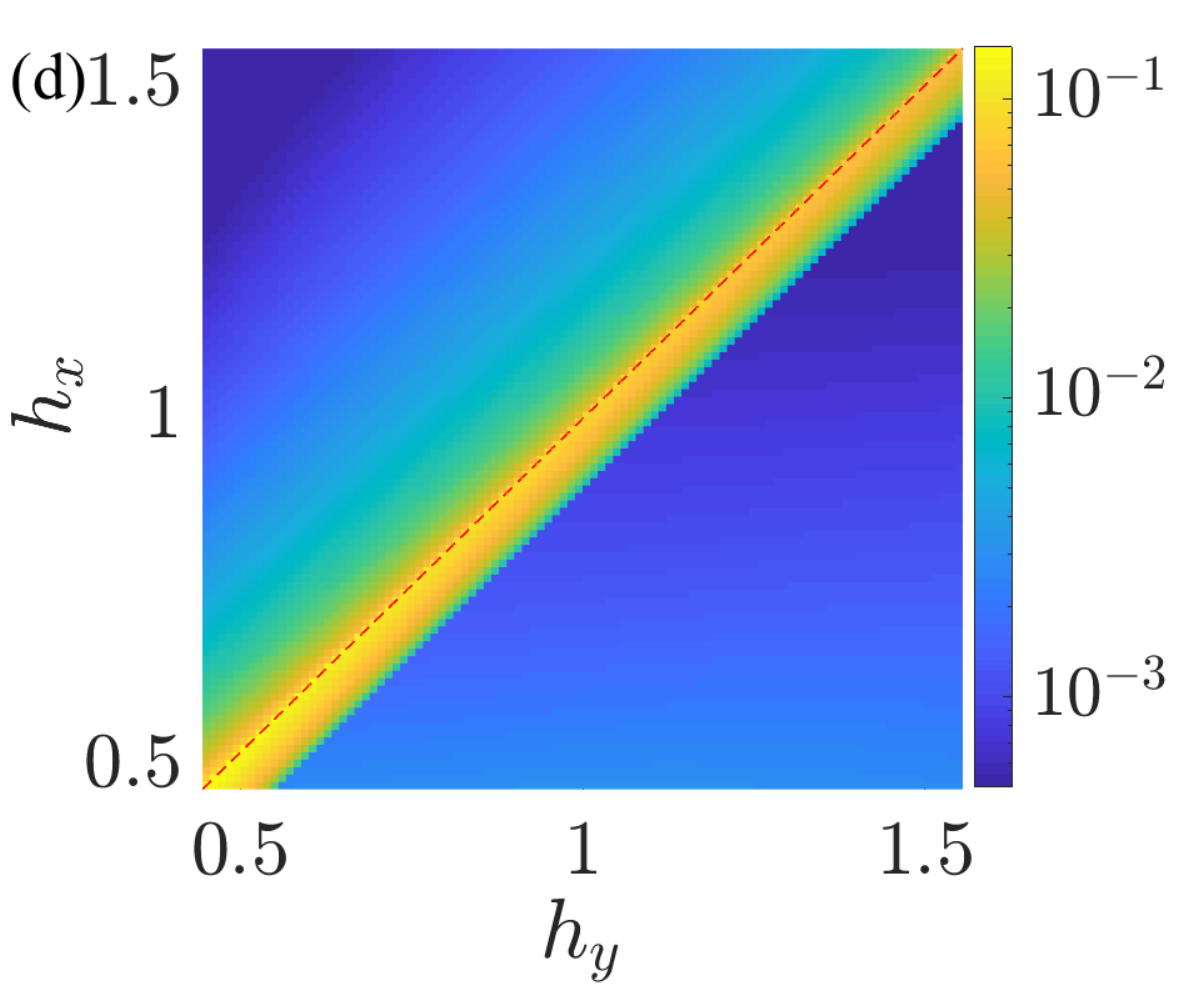}\caption{Ground-state susceptibility near EP for the non-Hermitian Ising
and Heisenberg model under a complex transverse field. (a--b)
Susceptibility $\chi(\boldsymbol{h},\boldsymbol{\delta})$ for
the non-Hermitian Ising model with $\delta_{x}=\delta_{y}$ (a)
and $\delta_{y}=0$ (b) for $N=8$. (c--d) Susceptibility for
the non-Hermitian Heisenberg model with $\delta_{x}=\delta_{y}$
(c) and $\delta_{y}=0$ (d) for $N=4$. All simulations use $J=1/2$
and $|\boldsymbol{\delta}|=\sqrt{\delta_{x}^{2}+\delta_{y}^{2}}=0.1$.}
\label{fig: state_susceptibility}
\end{figure}

In this section, we demonstrate how the non-Hermitian dynamics
dynamically selects a unique polarized environmental ground state.
To quantify the ground-state sensitivity, we define the ground-state
\emph{susceptibility} via the overlap between the perturbed and
unperturbed ground states: 
\begin{equation}
\chi(\boldsymbol{h},\boldsymbol{\delta})=-\ln|\langle G(\boldsymbol{h}+\boldsymbol{\delta})|G(\boldsymbol{h})\rangle|.\label{eq: susceptibility}
\end{equation}
where $\boldsymbol{h}\equiv(h_{x},h_{y})$ characterizes the
strength of the complex transverse field in the $(x,y)$ plane,
with both $h_{x}$ and $h_{y}$ being real numbers. We numerically
evaluate this susceptibility over a range of $(h_{x},h_{y})$
values for two coupling types: (i) $\delta_{x}=\delta_{y}$ and
(ii) $\delta_{y}=0$, using the same parameters as in Fig.~1.
As shown in Figs.~\ref{fig: state_susceptibility}(a,~c), when
$\delta_{x}=\delta_{y}$, the susceptibility $\chi$ is minimized
near the exceptional line (the locus of EPs in parameter space),
indicating that the ground state is particularly robust against
such perturbations. This observation explains the suppressed
$C[\rho(t)]$ decay in Figs.~1(b1,~c1). In contrast, for $\delta_{y}=0$,
Figs.~\ref{fig: state_susceptibility}(b,~d) show that $\chi$
increases as the system approaches the exceptional line from
the $h_{x}>h_{y}$ side. On the $h_{x}<h_{y}$ side, the susceptibility
decreases for the Heisenberg model but increases for the Ising
model.

\section{Quantum Circuit Protocol }

In this section, we show how the coherence $C[\rho(t)]$ can
be directly measured within a quantum circuit. With the initial
qubit state $\left(|0\rangle+|1\rangle\right)/\sqrt{2}$, the
total system--environment state is expressed as $|\psi(t)\rangle=\frac{1}{\sqrt{2}}\left(|0\rangle\otimes|\varphi_{0}(t)\rangle+|1\rangle\otimes|\varphi_{\boldsymbol{\delta}}(t)\rangle\right)$.
Substituting the qubit’s reduced density matrix $\rho(t)$ (obtained
by tracing out the environmental degrees of freedom) into the
$\ell_{1}$ norm of the coherence $C[\rho(t)]\equiv\sum_{i\neq j}|\rho_{ij}(t)|$
yields $C[\rho(t)]=\left|\langle\varphi_{0}(t)|\varphi_{\boldsymbol{\delta}}(t)\rangle/\langle\psi(t)|\psi(t)\rangle\right|$,
showing that the $\ell_{1}$ norm of the qubit’s coherence is
equivalent to the overlap between the two time-evolved environmental
states.

\subsection{Preparation of the initial state}

In this subsection, we detail the implementation of the operations
$U_{G}$ and $U_{G}^{-1}$ shown in Fig.~3(a). Specifically,
we begin by constructing the inverse operation $U_{G}^{-1}$,
which maps the ground state $|G\rangle_{\mathrm{E}}$ to the
initial product state $|0\rangle\otimes|0\rangle$. The unitary
operation $U_{G}$, which prepares the ground state, can then
be obtained by reversing this procedure. The inverse operation
$U_{G}^{-1}$ consists of three single-qubit rotation gates $R_{Y}(\theta_{1}),R_{Y}(\theta_{2}),R_{Y}(\theta_{3})$,
along with a CZ gate. To begin, the process starts from the ground
state $|G\rangle_{\mathrm{E}}\equiv\xi_{0}|00\rangle+\xi_{1}|01\rangle+\xi_{2}|10\rangle+\xi_{3}|11\rangle$.

First, the circuit begins by applying a rotation gate $R_{Y}(\theta_{1})$
to the second qubit, followed by a CZ gate applying on both qubits.
This sequence transforms the initial state $|G\rangle_{\mathrm{E}}$
into the intermediate state: 
\begin{align}
|\varphi_{1}\rangle & =U_{CZ}\left(I\otimes R_{Y}(\theta_{1})\right)|G\rangle_{\mathrm{E}},
\end{align}
where the rotation angle is defined as $\theta_{1}=2\arctan\!\left(\left(\xi_{2}-k\xi_{0}\right)/\left(\xi_{3}-k\xi_{1}\right)\right)$
and $k=-\operatorname{sign}(\xi_{0}\xi_{2}+\xi_{1}\xi_{3})\sqrt{\left(\xi_{2}^{2}+\xi_{3}^{2}\right)/\left(\xi_{0}^{2}+\xi_{1}^{2}\right)}$,
with $\operatorname{sign}(x)\equiv x/|x|$ for $x\neq0$.

The resulting state can be written as a tensor product: 
\[
|\varphi_{1}\rangle\equiv\left(a_{0}|0\rangle+a_{1}|1\rangle\right)\otimes\left(a_{2}|0\rangle+a_{3}|1\rangle\right).
\]
where the parameters are defined in terms of the original ground
state amplitudes $\xi_{i}$ : 
\begin{align}
a_{0} & =1/\sqrt{1+k^{2}},\\
a_{1} & =k/\sqrt{1+k^{2}},\\
a_{2} & =\sqrt{1+k^{2}}\left(-\xi_{1}\xi_{2}+\xi_{0}\xi_{3}\right)/\varsigma,\\
a_{3} & =\sqrt{1+k^{2}}\left[k\left(\xi_{0}\xi_{2}+\xi_{1}\xi_{3}\right)-\left(\xi_{2}^{2}+\xi_{3}^{2}\right)\right]/k\varsigma,
\end{align}
with the normalization factor $\varsigma\equiv\sqrt{\left(-k\xi_{0}+\xi_{2}\right)^{2}+\left(-k\xi_{1}+\xi_{3}\right)^{2}}$.
This transformation disentangles the ground state, yielding a
fully separable product form.

\begin{figure}
\includegraphics[width=1.7in]{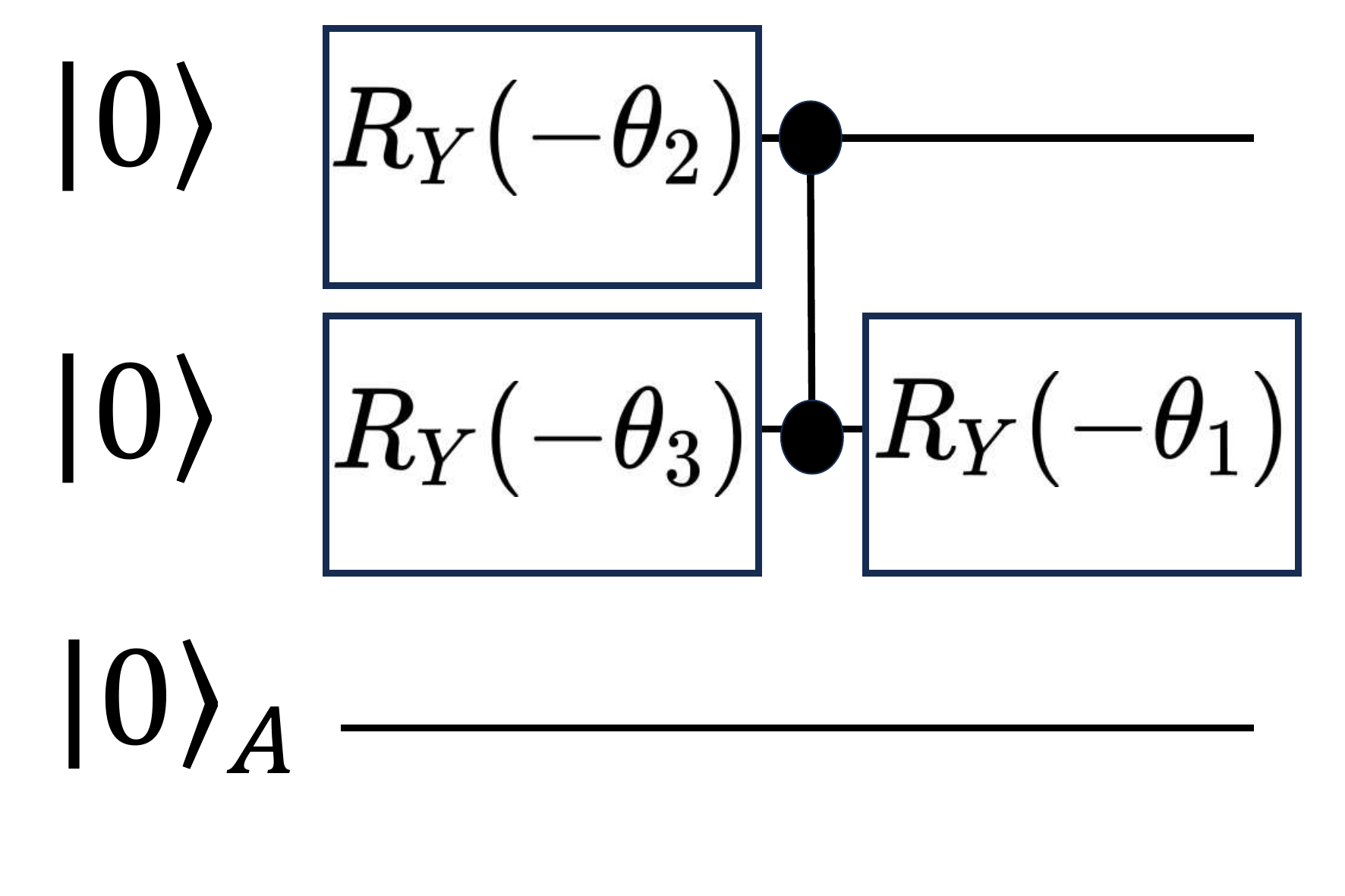}

\caption{Quantum circuit realization of $U_{G}$ in Fig.~3(a).}

\label{fig:QC_realization}
\end{figure}

Second, the rotation gate $R_{Y}(\theta_{2})$ is applied to
the first qubit, 
\begin{align}
|\varphi_{2}\rangle & =\left(R_{Y}(\theta_{2})\otimes I\right)|\varphi_{1}\rangle\equiv\gamma_{0}|00\rangle+\gamma_{1}|01\rangle,
\end{align}
where the amplitudes are given by 
\begin{align}
\gamma_{0}= & \frac{\left(a_{0}a_{3}+ka_{1}a_{3}\right)\left(-\xi_{1}\xi_{2}+\xi_{0}\xi_{3}\right)}{\tau},\\
\gamma_{1}= & \frac{a_{1}a_{3}\left(k\xi_{0}\xi_{2}-\xi_{2}^{2}+k\xi_{1}\xi_{3}-\xi_{3}^{2}\right)}{\tau}\nonumber \\
 & +\frac{a_{0}a_{3}\left(-k\xi_{0}^{2}+\xi_{0}\xi_{2}-k\xi_{1}^{2}+\xi_{1}\xi_{3}\right)}{\tau},
\end{align}
and the normalization factor is 
\begin{equation}
\tau\equiv\sqrt{\left(\left(a_{0}a_{3}\right)^{2}+\left(a_{1}a_{3}\right)^{2}\right)\left[\left(-k\xi_{0}+\xi_{2}\right)^{2}+\left(-k\xi_{1}+\xi_{3}\right)^{2}\right]}.
\end{equation}
The rotation angle is defined as $\theta_{2}=2\arctan\left(-a_{1}a_{3}/a_{0}a_{3}\right)$.
This step rotates the first qubit from the $X$-basis to the
$Z$-basis, mapping $|+\rangle=\left(|0\rangle+|1\rangle\right)/\sqrt{2}$
onto $|0\rangle$. The two-qubit state is projected into the
subspace spanned by $|00\rangle$ and $|01\rangle$, bringing
it closer to the target product state $|0\rangle\otimes|0\rangle$.

Finally, a rotation gate $R_{Y}(\theta_{3})$ is applied to the
second qubit, yielding the final state

\begin{align}
|00\rangle & =\left(I\otimes R_{Y}(\theta_{3})\right)|\varphi_{2}\rangle,
\end{align}
where the rotation angle is given by $\theta_{3}=2\arctan(-\gamma_{1}/\gamma_{0})$.
This final step completes the transformation from the ground
state $|G\rangle_{\mathrm{E}}$ to the fully separable state
$|0\rangle\otimes|0\rangle$.

The corresponding ground-state preparation can be achieved by
reversing the above procedure: the inverse sequence consists
of applying the gates in reverse order, with each rotation angle
negated. Specifically, the sequence begins with $R_{Y}(-\theta_{2})$
on the first qubit and $R_{Y}(-\theta_{3})$ on the second qubit.
Subsequently, a CZ gate is applied, and finally $R_{Y}(-\theta_{1})$
acts on the second qubit. This gate sequence realizes the unitary
transformation that prepares the ground state $|G\rangle_{\mathrm{E}}$,
as illustrated in Fig.~\ref{fig:QC_realization}. In the main
text, the ground state of the two-site non-Hermitian environment,
is represented in the quantum circuit by the normalized column
vector in the $Z$ basis $[(h_{x}+h_{y})/\beta,-\alpha/2\beta,-\alpha/2\beta,(h_{x}-h_{y})/\beta]^{T}$,
where $\alpha\equiv J-\sqrt{4h_{x}^{2}-4h_{y}^{2}+J^{2}}$ and
$\beta\equiv\sqrt{4h_{x}^{2}+J\alpha}$.

\subsection{Simulating the non-unitary dynamics}

This subsection demonstrates how the quantum circuit shown in
Fig.~3(a) simulates the non-unitary part of the dynamics in
the two-site non-Hermitian Ising Hamiltonian (Eq.~(1)), along
with the associated data postselection procedure~\cite{wang_2025_PRB}.
In the first time interval $dt$, the unitary operation $U_{G}$
prepares the ground state $|G\rangle_{\mathrm{E}}$ from the
initial product state $|0\otimes0\rangle$. This is followed
by unitary evolution under the operator $U(dt)$, resulting in
an intermediate state expressed in the eigenbasis of $Y$. To
fully implement the non-unitary evolution, an ancilla qubit is
introduced. In the limit of small $dt$, the non-unitary part
can be decomposed as 
\begin{align}
 & \exp\left(-\tilde{dt}\sum_{j=1}^{N}Y_{j}\right)\nonumber \\
\approx & \left(\frac{\cos\tilde{dt}}{\sqrt{2}}\right)^{-N}\prod_{j=1}^{N}\left(\frac{\cos\tilde{dt}}{\sqrt{2}}-\frac{\sin\tilde{dt}}{\sqrt{2}}P_{j,\uparrow_{y}}+\frac{\sin\tilde{dt}}{\sqrt{2}}P_{j,\downarrow_{y}}\right),
\end{align}
where $\tilde{dt}=\left(h_{y}+\delta_{y}\right)dt$, $P_{j,\uparrow_{y}}\equiv\left(1+Y_{j}\right)/2$,
$P_{j,\downarrow_{y}}\equiv\left(1-Y_{j}\right)/2$. Each non-unitary
term $\left[\left(\cos\tilde{dt}/\sqrt{2}\right)-\left(\sin\tilde{dt}/\sqrt{2}\right)P_{j,\uparrow_{y}}+\left(\sin\tilde{dt}/\sqrt{2}\right)P_{j,\downarrow_{y}}\right]$
in the product is implemented via a sequence of quantum operations
involving an ancilla qubit, which enables the effective non-unitary
evolution through post-selection. The simulation proceeds as
follows:

First, the ancilla qubit is initialized to the state 
\begin{align}
 & e^{-i\left(\pi/4+\tilde{dt}\right)Y}|0\rangle_{A}\nonumber \\
 & =\frac{1}{\sqrt{2}}\left(\cos\tilde{dt}-\sin\tilde{dt}\right)|0\rangle_{A}+\frac{1}{\sqrt{2}}\left(\cos\tilde{dt}+\sin\tilde{dt}\right)|1\rangle_{A}.
\end{align}
Second, unitary rotations $R\equiv(e^{i\pi/4}I+e^{-i\pi/4}\sum_{a=X,Y,Z}a)/2$
are applied to the first system qubit, such that $R|\uparrow_{y}\rangle(|\downarrow_{y}\rangle)=|\uparrow\rangle(|\downarrow\rangle)$,
resulting in state $\sum_{Y_{1},Z_{2}}c_{Y_{1},Z_{2}}(R|Y_{1}\rangle)\otimes|Z_{2}\rangle.$
Third, a controlled-NOT gate is applied between the first system
qubit (control) and the ancilla qubit (target), resulting in
the state: 
\begin{align}
\sum_{Z_{2}}c_{\uparrow_{y},Z_{2}}(R| & \uparrow_{y}\rangle)\otimes|Z_{2}\rangle\otimes\left(e^{-i\left(\pi/4+\tilde{dt}\right)Y}|0\rangle_{A}\right)\nonumber \\
+\sum_{Z_{2}}c_{\downarrow_{y},Z_{2}}(R| & \downarrow_{y}\rangle)\otimes|Z_{2}\rangle\otimes\left(Xe^{-i\left(\pi/4+\tilde{dt}\right)Y}|0\rangle_{A}\right).
\end{align}
Fourth, the ancilla qubit is measured, and only the outcomes
corresponding to $|0\rangle_{a}$ are retained. The resulting
post-selected state becomes: 
\begin{align}
\sum_{Z_{2}}c_{\uparrow_{y},Z_{2}}(R| & \uparrow_{y}\rangle)\otimes|Z_{2}\rangle\otimes\frac{\cos\tilde{dt}-\sin\tilde{dt}}{\sqrt{2}}|0\rangle_{A}\nonumber \\
+\sum_{Z_{2}}c_{\downarrow_{y},Z_{2}}(R| & \downarrow_{y}\rangle)\otimes|Z_{2}\rangle\otimes\frac{\cos\tilde{dt}+\sin\tilde{dt}}{\sqrt{2}}|0\rangle_{A}.
\end{align}
Fifth, the inverse rotation $R^{-1}$ is applied to the first
qubit, and the ancilla qubit is discarded. This operation effectively
implements the action of $\left[\left(\cos\tilde{dt}/\sqrt{2}\right)-\left(\sin\tilde{dt}/\sqrt{2}\right)P_{j,\uparrow_{y}}+\left(\sin\tilde{dt}/\sqrt{2}\right)P_{j,\downarrow_{y}}\right]$
on the left of the state $\sum_{Y_{1},Z_{2}}c_{Y_{1},Z_{2}}|Y_{1}\rangle\otimes|Z_{2}\rangle$.
The same procedure is subsequently applied to the second qubit
($N=2$). Upon measurement of the ancilla qubit in the $|0\rangle_{a}$
state for both sites, the final system state becomes: 
\begin{equation}
\left(\frac{\cos\tilde{dt}}{\sqrt{2}}\right)^{2}\exp(-idtH(\boldsymbol{h},\boldsymbol{\delta}))|G\rangle_{\mathrm{E}}
\end{equation}
where the two-site Hamiltonian is $H(\boldsymbol{h},\boldsymbol{\delta})=-JZ_{1}Z_{2}-\left(h_{x}+\delta_{x}\right)\left(X_{1}+X_{2}\right)-i\left(h_{y}+\delta_{y}\right)\left(Y_{1}+Y_{2}\right)$
and the implemented operation results in a deterministic overall
amplitude factor $\left(\cos\tilde{dt}/\sqrt{2}\right)^{2}$.
To compensate for this factor and recover the target non-Hermitian
dynamical evolution, the corresponding factor $\left(\cos\tilde{dt}/\sqrt{2}\right)^{-2}$
is calculated and applied to the measured data in the subsequent
analysis.

To evaluate the $C[\rho(t)]$ in the two-site non-Hermitian Ising
Hamiltonian (Eq.~(1)), both the overlap $\left|\langle\varphi_{0}(dt)|\varphi_{\boldsymbol{\delta}}(dt)\rangle\right|$
and the normalization factor $\langle\psi(t)|\psi(t)\rangle$
must be determined. The analysis proceeds in two steps. First,
the normalization factor is given by 
\begin{align}
 & \langle\psi(dt)|\psi(dt)\rangle\nonumber \\
 & =\frac{1}{2}\left(\langle\varphi_{\boldsymbol{\delta}}(dt)|\varphi_{\boldsymbol{\delta}}(dt)\rangle+\langle\varphi_{0}(dt)|\varphi_{0}(dt)\rangle\right)\nonumber \\
 & =\frac{1}{2}\left(_{\mathrm{E}}\langle G|e^{idtH^{\dagger}(\boldsymbol{h},\boldsymbol{\delta})}e^{-idtH(\boldsymbol{h},\boldsymbol{\delta})}|G\rangle_{\mathrm{E}}+1\right)\nonumber \\
 & =\frac{1}{2}\left(\frac{\cos\tilde{dt}}{\sqrt{2}}\right)^{-4}\left(\sum_{j=0}^{3}\left|c_{j}\right|^{2}\right)+\frac{1}{2},
\end{align}
where the evolved state $e^{-idtH(\boldsymbol{h},\boldsymbol{\delta})}|G\rangle_{\mathrm{E}}\equiv c_{0}|00\rangle+c_{1}|01\rangle+c_{2}|10\rangle+c_{3}|11\rangle$
is obtained by implementing both the unitary and non-unitary
parts of the evolution operator $e^{-idtH(\boldsymbol{h},\boldsymbol{\delta})}$
within a quantum circuit. The factor $\left(\cos\tilde{dt}/\sqrt{2}\right)^{-4}$
derived from the non-unitary evolution contribution, is multiplied
by the measured data according to the analytical formula to obtain
the normalization factor.

Second, the overlap $\left|\langle\varphi_{0}(dt)|\varphi_{\boldsymbol{\delta}}(dt)\rangle\right|$
is simplified as: 
\begin{align}
 & \left|\langle\varphi_{0}(dt)|\varphi_{\boldsymbol{\delta}}(dt)\rangle\right|\nonumber \\
 & =\left|_{\mathrm{E}}\langle G|e^{idtH^{\dagger}(\boldsymbol{h},0)}e^{-idtH(\boldsymbol{h},\boldsymbol{\delta})}|G\rangle_{\mathrm{E}}\right|\nonumber \\
 & =\left|_{\mathrm{E}}\langle G|e^{-idtH(\boldsymbol{h},\boldsymbol{\delta})}|G\rangle_{\mathrm{E}}\right|\nonumber \\
 & =\left|\langle00|U_{G}^{-1}e^{-idtH(\boldsymbol{h},\boldsymbol{\delta})}U_{G}|00\rangle\right|\nonumber \\
 & =\left[\frac{\cos\left((h_{y}+\delta_{y})dt\right)}{\sqrt{2}}\right]^{-2}\left|c_{0}^{\prime}\right|,
\end{align}
where $c_{0}^{\prime}$ is the amplitude of the $|00\rangle$
component in the transformed state $U_{G}^{-1}e^{-idtH(\boldsymbol{h},\boldsymbol{\delta})}U_{G}|00\rangle$.
In this representation, $U_{G}$ prepares the ground state from
$|00\rangle$, and its inverse projects the evolved state back
onto $|00\rangle$, allowing direct extraction of the $C[\rho(t)]$.
The final overlap value is obtained by multiplying the measured
amplitude by the known factor $\left(\cos\tilde{dt}/\sqrt{2}\right)^{-2}$.

\subsection{Adaptive-step trotterization scheme}

In practice, imperfections in current quantum hardware introduce
a trade-off: smaller time steps improve accuracy but necessitate
deeper circuits, thereby amplifying noise. In this subsection,
we adopt an adaptive-step trotterization scheme~\cite{Zhao_2023_PRX,Zhao_2024_PRL}
to balance these competing effects; it dynamically adjusts the
step size $dt$ to ensure the expectation value of the target
observable remains within a predefined bound.

To employ adaptive-step trotterization scheme, we examine a two-site
system ($N=2$) and determine the maximum allowable variable
step size $dt$. Specifically, we simulate the time evolution
using IBM Qiskit~\cite{cross_2018_ibm}, adjusting $dt$ so
that the deviation between the Trotterized and exact (non-Trotterized)
evolutions remains below a tolerance of $10^{-2}$. As shown
in Fig.~3(b), the quantum circuit data of $C[\rho(t)]$ obtained
under this condition demonstrate that the adaptive-step protocol
effectively maintains accuracy while minimizing circuit depth.
\end{document}